\documentclass[12pt]{article}
\usepackage{booktabs}
\usepackage{caption} 
\usepackage{subcaption} 
\usepackage{amsfonts}
\usepackage{amsmath}
\usepackage{amssymb}
\usepackage{bm}
\usepackage{setspace}
\usepackage{graphicx,psfrag,epsf}
\usepackage{natbib}
\usepackage{xcolor}
\usepackage{calrsfs}
\usepackage{tabularx}
\usepackage{ragged2e}
\definecolor{winered}{rgb}{0.5,0,0}
\usepackage[bookmarks=true, bookmarksnumbered=true, allbordercolors={1 1 1}]{hyperref}
\hypersetup{
  colorlinks   = true, 
  urlcolor     = blue, 
  linkcolor    = blue, 
  citecolor    = winered,
}
\usepackage[open=true, numbered=true]{bookmark}
\usepackage{float}
\usepackage[toc,page]{appendix}
\usepackage{scalefnt}
\usepackage{afterpage}
\usepackage[left]{lineno}
\usepackage{varioref}
\usepackage{cleveref}
\usepackage{algorithm}
\usepackage[end]{algpseudocode}
\algnewcommand\algorithmicto{\textbf{to }}
\usepackage{threeparttablex}
\usepackage{xurl}
\usepackage{mathptmx} 
\usepackage[T1]{fontenc}
\usepackage{enumitem}
\usepackage{hyperref}
\usepackage{xargs}                      
\usepackage[colorinlistoftodos,prependcaption,textsize=tiny]{todonotes}
\newcommandx{\unsure}[2][1=]{\todo[linecolor=red,backgroundcolor=red!25,bordercolor=red,#1]{#2}}
\newcommandx{\change}[2][1=]{\todo[linecolor=blue,backgroundcolor=blue!25,bordercolor=blue,#1]{#2}}
\newcommandx{\info}[2][1=]{\todo[linecolor=OliveGreen,backgroundcolor=OliveGreen!25,bordercolor=OliveGreen,#1]{#2}}
\newcommandx{\improvement}[2][1=]{\todo[linecolor=Plum,backgroundcolor=Plum!25,bordercolor=Plum,#1]{#2}}
\newcommandx{\thiswillnotshow}[2][1=]{\todo[disable,#1]{#2}}

\makeatletter
\def\blfootnote{\xdef\@thefnmark{}\@footnotetext}
\makeatother

\newcommand{\blind}{0}

\begin{document}

\def\spacingset#1{\renewcommand{\baselinestretch}%
{#1}\small\normalsize} \spacingset{1}

\defcitealias{Chenetal2021}{CDG}

\if0\blind
{
  \title{\bf Probabilistic Quantile Factor Analysis\blfootnote{The authors gratefully acknowledge helpful comments from participants of the 2023 SNDE symposium and the IAAE 2023 in Oslo.
This paper should not be reported as representing the views of Norges Bank. The views expressed are those of the authors and do not necessarily reflect those of Norges Bank. The authors report there are no competing interests to declare.}}
  \author{Dimitris Korobilis\thanks{\href{mailto:Dimitris.Korobilis@glasgow.ac.uk}{Dimitris.Korobilis@glasgow.ac.uk}; \href{mailto:Dimitris.Korobilis@bi.no}{Dimitris.Korobilis@bi.no}}\hspace{.2cm}\\
University of Glasgow,\\
BI Norwegian Business School    \\
    and \\
    Maximilian Schr\"oder\thanks{\href{mailto:maximilian.schroder@bi.no}{Maximilian.Schroder@bi.no}} \\
BI Norwegian Business School \\
    Norges Bank}
  \maketitle
} \fi

\if1\blind
{
  \bigskip
  \bigskip
  \bigskip
  \begin{center}
    {\LARGE\bf Title}
\end{center}
  \medskip
} \fi

\bigskip

\maketitle
\begin{abstract}
\noindent This paper extends quantile factor analysis to a probabilistic variant that incorporates regularization and computationally efficient variational approximations. We establish through synthetic and real data experiments that the proposed estimator can, in many cases, achieve better accuracy than a recently proposed loss-based estimator. We contribute to the factor analysis literature by extracting new indexes of \emph{low}, \emph{medium}, and \emph{high} economic policy uncertainty, as well as \emph{loose}, \emph{median}, and \emph{tight} financial conditions. We show that the high uncertainty and tight financial conditions indexes have superior predictive ability for various measures of economic activity. In a high-dimensional exercise involving about 1000 daily financial series, we find that quantile factors also provide superior out-of-sample information compared to mean or median factors.
\bigskip

\noindent \emph{Keywords:} variational Bayes; penalized factors; quantile regression
\bigskip 

\noindent \emph{JEL Classification:}\ C11, C31, C32, C52, C53
\end{abstract}
\thispagestyle{empty} 

\newpage

\setcounter{page}{1}
\spacingset{1.8} 

\section{Introduction}
A fundamental contribution of \cite{SargentSims1977} was the illustration that two unobserved factors can be successful in summarizing the information in a large number of macroeconomic variables, an approach they coined ``measurement without theory''. Since this early application of factor analysis in economics, empirical investigations relying on this class of models have evolved tremendously,\footnote{For example, the factor model has become the ground for characterizing international business cycle comovements \citep{Koseetal2003}; for modeling mixed frequency macro data \citep{MarianoMurasawa2003}; and for structural vector autoregressive analysis \citep{Bernankeetal2005}, among numerous other applications.} as has our understanding of various statistical estimation approaches to factor models and their properties.\footnote{See for instance, \cite{StockWatson2002a}.} During the prolonged period of stability in macroeconomic volatility known as the Great Moderation (ca. 1982–2007), constant parameter factor models have experienced good fit. After the Great Recession shock of 2007–2009, factor models that feature structural breaks, stochastic volatility, and other flexible extensions have emerged in the literature.\footnote{\cite{Batesetal2013,KoopKorobilis2014}.} More recently, several authors suggest that modeling each quantile of the distribution of time series data -- a statistical procedure widely known as quantile regression \citep{Koenker2005} -- can be more beneficial for inference and forecasting.\footnote{Most notably this point was made in \cite{Adrianetal2019}, but see also \cite{Iacopinietal2022}, \cite{Korobilis2017} and \cite{LopezLoria2019}.} The argument in favor of quantile models is that different predictor variables and model features may be relevant for explaining each quantile of a series' distribution. Therefore, it is no surprise that papers such as \cite{AndoBai2020}, \cite{Chenetal2021} (hereafter \citetalias{Chenetal2021}) and \cite{Clarketal2021} attempt to estimate quantile factor models that deal with the econometric problem of allowing different factors to characterize each quantile of $n$ comoving economic variables observed over $T$ time periods.

The purpose of this paper is to contribute to this emerging literature in econometrics, by proposing a new probabilistic quantile factor analysis framework. Our motivation and starting point is the specification of \citetalias{Chenetal2021}, which provides the benchmark for characterizing complex, asymmetric distributions of unobserved factors. For a $T\times n$ panel $\mathbf{x}$ of $n$ variables observed over $T$ time periods, the quantile factor model is of the form $\mathbf{x} = \bm f_{(\tau)} \bm \lambda_{(\tau)}^{\prime} + \bm u_{(\tau)}$, where $\bm f_{(\tau)}$ is the $T \times r$ matrix of factors, $r \ll n$, $\bm \lambda_{(\tau)}$ is an $n \times r$ loadings matrix, and $\tau \in (0,1)$ denotes the quantile level. \citetalias{Chenetal2021} propose an iterative procedure that can be viewed as a generalization of principal component analysis (PCA) to the case of a check function loss (while PCA minimizes a quadratic loss function). These authors allow for a different number of factors to affect different quantiles by means of testing using information criteria. We argue that working explicitly in a probabilistic setting naturally incorporates likelihood penalization, a crucial feature for accurate estimation and regularization of extreme quantiles. In monthly or quarterly macroeconomic time series, for example, extreme quantiles may correspond to a fairly small proportion of the observed sample, making the unrestricted estimation of quantile factors imprecise.

In order to deal with estimation in a computationally efficient manner, we build on the statistical machine learning literature and derive a variational Bayes algorithm for a parametric quantile factor analysis model based on an asymmetric Laplace likelihood. This contribution is an extension of the factor analysis model of \cite{BealGhahramani1999}, and provides a much faster alternative to the flexible probabilistic model of \cite{AndoBai2020} that relies on computationally intensive Markov chain Monte Carlo (MCMC) methods.\footnote{See \cite{Bleietal2017} for a review of the properties and benefits of variational Bayes inference.} In order to regularize the likelihood, the sparse Bayesian learning prior of \cite{Tipping2001} is adopted, which is found to provide excellent numerical performance. Both features make the proposed parametric framework especially suitable for large-scale applications and real-time empirical investigation that can inform macroeconomic policy decisions. 

We evaluate the numerical performance of the variational algorithm using synthetic data experiments. The benchmark simulation design features flexible disturbance terms, ranging from heavy-tailed to skewed bimodal distributions with and without correlation. By performing numerical comparisons to the loss-based quantile factor estimator of \citetalias{Chenetal2021}, it is established that the probabilistic quantile factor methodology recovers at least as accurately, in the vast majority of cases, the true factors, especially in the tails of the distribution. Furthermore, the probabilistic estimator with shrinkage prior is more robust to changes in the signal-to-noise ratio and the size of the information set, whereas the loss-based estimator's performance declines as the number of variables increases. We support these results with additional simulation studies in which we compare the performance of our probabilistic estimator to an MCMC algorithm. The results suggest that approximation error is negligible for a large set of different prior distributions and highlight the applicability of our proposed framework to large-scale empirical settings.

We illustrate how the proposed methodology opens up new avenues for empirical research, by means of a novel approach to extracting macroeconomic indexes. In particular, we conduct three empirical exercises using diverse datasets spanning different concepts and observed frequencies. In the first exercise, we extract quantile financial conditions indexes (FCIs), at the 10th, 50th, and 90th percentile levels, to forecast economic conditions in the U.S. at weekly frequency. The data we use are the 105 components of the popular National FCI (NFCI) maintained by the Federal Reserve Bank of Chicago. We interpret the three estimated quantile factors from the 105 series as indexes of ``loose'', ``median'' and ``tight'' financial conditions. We find that the ``tight FCI'' extracted with the proposed estimator improves upon the forecast performance of the aggregate NFCI, as well as principal components and loss-based quantile factor estimation. This result suggests that linear combinations (factors) of the 90th percentiles of financial variables provide useful information on the economic outlook that is not contained in mean or median factors. In the second exercise, quantile factors are extracted from nine disaggregated categories that comprise the (aggregate) economic policy uncertainty (EPU) index for the U.S. developed by \cite{Bakeretal2016}. These individual categories of uncertainty (e.g. monetary policy, fiscal policy, regulation) are utilized in order to approximate the full distribution of aggregate economic policy uncertainty by means of quantile factors. We interpret the 10th, 50th and 90th percentile factors, as ``low'', ``medium'' and ``high'' uncertainty indexes. Compared to the aggregate EPU index, we find that these quantile-specific indexes provide forecast performance gains for industrial production and the interest rate, while their performance is on par for inflation. In the case of industrial production, the 90th percentile ``high uncertainty'' factor is the driver behind the performance gains, suggesting that industrial production forecasts benefit from capturing spells of high uncertainty more accurately. In the final exercise, we extract quantile factors from a large financial data set at daily frequency that covers about 1000 series, to explore the model's favourable properties in a large-scale empirical setting. Subsequently, we test the performance of the factors in forecasting quarterly U.S. real GDP growth using mixed-data sampling regressions, following \cite{andreou2013should}. In line with the first exercise, we find that the 90th percentile factor extracted with our model is particularly informative for the real economic outlook, providing non-negligible forecast performance gains over alternative estimators.

We organize the remainder of this paper as follows: Section 2 introduces the probabilistic quantile factor analysis framework and the main steps of the variational Bayes estimation algorithm. It also explains how the basic setting could be expanded. In Section 3, we illustrate the numerical properties of the proposed algorithm using simulated data generated from factor structures with very flexible distributions. We dedicate Section 4 to demonstrating our estimator in a variety of empirical settings and datasets. Section 5 concludes the paper.

\section{Methodology}
Let $x_{it}$ denote variable $i=1,\dots, n$ observed over period $t=1,\dots, T$. Our starting point is a factor model decomposition, for each conditional quantile $\tau \in (0,1)$ of $x_{it}$, of the form
\begin{equation}
Q_{(\tau)}\left( x_{it} \vert \bm \lambda_{i,(\tau)}, \bm f_{t,(\tau)} \right)  = \bm \lambda_{i,(\tau)}^{\prime} \bm f_{t,(\tau)},
\end{equation}
where $\bm \lambda_{i,(\tau)}$ and $\bm f_{t,(\tau)}$ are $r \times 1$ vectors with $r \ll n$ the number of unobserved factors that is allowed to vary across quantile levels.\footnote{As in \citetalias{Chenetal2021} we can write $r=r_{(\tau)}$ to denote this feature, but we don't do so for notational simplicity. In this section we assume that all quantiles have the same number of factors, but the regularization features of our likelihood-based approach mean that empirically different numbers of factors may affect each quantile of $x_{it}$.} The above equation admits the following regression form
\begin{equation}
x_{it} =  \bm \lambda_{i,(\tau)}^{\prime} \bm f_{t,(\tau)} + u_{it,(\tau)},
\end{equation}
where $u_{it,(\tau)}$ is a quantile-dependent idiosyncratic error. In line with the exact factor model assumptions the $u_{it,(\tau)}$ are uncorrelated across variables $i$, implying that only the common component $\bm \lambda_{i,(\tau)}^{\prime} \bm f_{t,(\tau)}$ captures comovements of the variables $x_{it}$, at each quantile level $\tau$.

In the factor analysis model for the conditional mean of $ x_{it}$, estimation can be implemented either by maximizing the likelihood \citep{Bishop1999,BealGhahramani1999} or by minimizing a quadratic loss function \citep{StockWatson2002a}.\footnote{The latter approach results in principal component estimates of $\bm f_{t,(\tau)}$ and least squares estimates of $\bm \lambda_{i,(\tau)}$.} When moving to quantile factor models, \cite{Chenetal2021} propose to obtain the optimal parameter values for $\bm \lambda_{i,(\tau)}$ and $\bm f_{t,(\tau)}$ by minimizing the expected loss of $u$ under the check function $\rho_{\tau}(u) = u_{it,(\tau)} \left(\tau - \mathbb{I}\lbrace u_{it,(\tau)} \leq 0 \rbrace \right)$, where $ \mathbb{I}$ is the indicator function \citep[see][pp. 5-6]{Koenker2005}. In this paper, we propose a variational Bayes estimation algorithm for probabilistic quantile factor models that builds on the legacy of factor analysis models as utilized in the statistical machine learning literature.

\subsection{ \textit{Likelihood-based Representation and Priors}}
A probabilistic approach to quantile factor analysis requires to replace loss-based estimation with an equivalent likelihood-based problem. Similar to least squares estimation being equivalent to regression with a Gaussian error term, it is well established \citep{YuMoyeed2001} that quantile regression estimates can be recovered as the maximum of the following asymmetric Laplace likelihood
\begin{equation}
AL_{(\tau)}(u_{it,(\tau)} ; 0,\sigma_{i,(\tau)}) = \frac{\tau(\tau-1)}{\sigma_{i,(\tau)}}\exp \left \lbrace -\frac{1}{\sigma_{i,(\tau)}}\rho_{\tau} \left( u_{it,(\tau)} \right) \right \rbrace,
\end{equation}
where $\sigma_{i,(\tau)}$ is a scale parameter and $\rho_{\tau}(\bullet)$ is the check function defined previously. The asymmetric Laplace distribution admits a location-scale mixture of normals representation, such that the $AL$ distributed $u$ can equivalently be written as
\begin{equation}
u_{it,(\tau)} = \frac{1-2\tau}{1-\tau} z_{it,(\tau)} +  \sqrt{\frac{2 \sigma_{i,(\tau)} z_{it,(\tau)}}{\tau \left( 1- \tau \right)}}v_{it},
\end{equation}
where $z_{it,(\tau)} \sim Exp(\sigma_{i,(\tau})$ and $v_{it} \sim N(0,1)$. By defining the quantities $ \bm z_{i,(\tau)} = \left[ z_{i1,(\tau)},...,z_{iT,(\tau)} \right]^{\prime}$, $\bm z_{(\tau)} = \left[\bm z_{1,(\tau)},...,\bm z_{n,(\tau)} \right]$, $\bm \sigma_{(\tau)}= \left[\sigma_{1,(\tau)},...,\sigma_{n,(\tau)} \right]$ and $\bm f_{(\tau)} = \left[ \bm f_{1,(\tau)},...,\bm f_{T,(\tau)} \right]^{\prime}$ the likelihood function of the quantile factor model -- conditional on $\bm z$ and the quantile level $\tau$ -- can thus be written as
\begin{equation}
p \left( \mathbf{x} \vert \bm \lambda_{(\tau)}, \bm f_{(\tau)} , \bm z_{(\tau)}, \bm \sigma_{(\tau)} \right) = \prod_{i=1}^{n} \prod_{t=1}^{T} N \left( \bm \lambda_{i,(\tau)}^{\prime} \bm f_{t,(\tau)}  + \sigma_{i,(\tau)} \frac{1-2\tau}{1-\tau} \bm z_{it,(\tau)}, \frac{2\sigma_{i,(\tau)}}{\tau(1-\tau)} \bm z_{it,(\tau)} \right). \label{ALQF_lik}
\end{equation}
Given that this transformed likelihood is conditionally normal and independent for each quantile level $\tau$, inference is markedly simplified.

One practical benefit of likelihood-based analysis of the quantile factor model is that regularization is easily incorporated via a penalized likelihood. Following \cite{Bishop1999} and others we regularize the likelihood by means of Bayesian prior distributions. In particular, for the loadings matrix we adopt the sparse Bayesian learning prior of \cite{Tipping2001} which takes the form
\begin{equation}
\bm \lambda_{(\tau)}\vert \bm \alpha_{(\tau)} \sim \prod_{i=1}^{n}\prod_{j=1}^{r} N \left( 0, \alpha_{ij,(\tau)}^{-1} \right), \text{ \ \ } \bm \alpha_{(\tau)} \sim \prod_{i=1}^{n}\prod_{j=1}^{r} G \left(a_0,b_0\right),
\end{equation}
where $\bm \alpha_{(\tau)} = \left[\alpha_{11,(\tau)},...,\alpha_{ij,(\tau)}\right]$. In practice, in all our calculations using synthetic and real data, we follow \cite{Tipping2001} and set $a_0=b_0=0.0001$ such that the gamma distribution approximates the uniform distribution. The scale parameter has prior $\bm \sigma_{(\tau)} \sim \prod_{i=1}^{n} G ^{-1} (r_0,s_0)$. The remaining two latent variables, namely $\bm f_{(\tau)}$ and $\bm z$, by definition have priors $\bm f_{(\tau)}  \sim \prod_{t=1}^{T} N _{r} \left(\bm 0_{r \times 1}, \bm I_{r} \right)$ and $\bm z_{(\tau)} \sim \prod_{i=1}^{n}\prod_{t=1}^{T} Exp(1)$, respectively. Therefore, the joint prior can be factorized as follows
\begin{equation}
p \left ( \bm \lambda_{(\tau)},\bm f_{(\tau)},\bm z_{(\tau)}, \bm \sigma_{(\tau)},\bm \alpha_{(\tau)} \right)  =  p \left ( \bm \lambda_{(\tau)} \vert \bm \alpha_{(\tau)} \right) p \left(\bm \alpha_{(\tau)}\right)  p \left(\bm f_{(\tau)}\right) p \left(\bm z_{(\tau)}\right)  p \left(\bm \sigma_{(\tau)}\right),
\end{equation}
where the individual densities are provided above.

\subsection{ \textit{Variational Bayes Inference}}

For a family of tractable densities $q\left( \bm \theta_{(\tau)} \right) \in \mathcal{Q}$, we want to find a density $q^{\star}$ that best approximates the intractable posterior $p(\bm \theta_{(\tau)} \vert \mathbf{x})$, where $\bm \theta_{(\tau)} = \left( \bm \lambda_{(\tau)}, \bm f_{(\tau)}, \bm z_{(\tau)}, \bm \sigma_{(\tau)} \right)$ are the model parameters and latent variables.\footnote{For notational simplicity, in this subsection we ignore the parameter $\bm \alpha_{(\tau)}$ that shows up only in the prior of $\bm \lambda_{(\tau)}$ and focus on parameters that show up in the likelihood function. That is, we proceed as if $\bm \alpha_{(\tau)}$ was fixed/known, a case that would result in a ridge-regression posterior median for $\bm \lambda_{(\tau)}$. The Appendix shows full derivations under the sparse Bayesian learning prior, i.e. when $\bm \alpha_{(\tau)}$ is a random variable.} That is, we seek to minimize the following loss
\begin{equation}
q^{\star} \left( \bm \theta_{(\tau)} \right) = \arg \min_{q \in \mathcal{Q}} \mathbb{D}_{KL} \left(q\left( \bm \theta_{(\tau)} \right)  \vert \vert  p(\bm \theta_{(\tau)} \vert \mathbf{x}) \right),
\end{equation}
where $\mathbb{D}_{KL}=\int_{\bm \Theta} q\left( \bm \theta_{(\tau)} \right) \log\left \lbrace \frac{q\left( \bm \theta_{(\tau)} \right)}{p\left( \bm \theta_{(\tau)} \vert \mathbf{x} \right)} \right \rbrace d\bm \theta$ denotes the Kullbach-Leibler (KL) divergence and $\bm \Theta$ is the support of $\bm \theta$.\footnote{The KL divergence is an information-theoretic measure of promixity of densities. However, notice that the densities $q$ and $p$ are defined in different spaces, such that the measure $\mathbb{D}_{KL}$ is not a true KL divergence.} Given the need to optimize over a family of distribution functions $q$, the solution to this problem requires the application of variational calculus. By using the definition of the KL divergence and Bayes theorem, the quantity $\mathbb{D}_{KL}$ above can be written as in \citep{Bleietal2017}
\begin{eqnarray}
\mathbb{D}_{KL} \left(q\left( \bm \theta_{(\tau)} \right)  \vert \vert  p(\bm \theta_{(\tau)} \vert \mathbf{x}) \right) & =  & \mathbb{E}\left[ \log q \left( \bm \theta_{(\tau)} \right) \right] - \mathbb{E}\left[ \log p \left( \bm \theta_{(\tau)} \vert \mathbf{x} \right) \right],   \\
 & =  & \mathbb{E}\left[ \log q \left( \bm \theta_{(\tau)} \right) \right] - \mathbb{E} \left[\log p \left( \mathbf{x} \vert \bm \theta_{(\tau)}  \right) \right] - \nonumber \\
 &&\mathbb{E} \left[ \log p \left( \bm \theta_{(\tau)}  \right) \right] + \mathbb{E} \left[ \log p \left( \mathbf{x}  \right) \right], \label{VBdecomp}
\end{eqnarray}
where all expectations are w.r.t. the variational density $q\left( \bm \theta_{(\tau)} \right)$. The last term is the marginal likelihood and it can neither be calculated analytically, nor does it involve the parameters of interest $\bm \theta_{(\tau)}$. Therefore, it is sufficient to minimize the first three terms on the right-hand side of equation \eqref{VBdecomp}, or equivalently maximize their negative value, which is known as the evidence lower bound (ELBO)
\begin{equation}
ELBO =  \mathbb{E} \left[ \log p \left( \mathbf{x} \vert \bm \theta_{(\tau)}  \right) \right] + \mathbb{E} \left[ \log p \left( \bm \theta_{(\tau)}  \right) \right] - \mathbb{E}\left[ \log q \left( \bm \theta_{(\tau)} \right) \right].
\end{equation}
From this point, optimization is typically simplified by factorizing the variational posterior into groups of independent densities, an assumption known in physics as mean-field inference. Different factorizations are possible, but for the quantile factor model we follow \cite{Limetal2020} and \cite{Bishop1999} and assume the following partitioning of the joint variational posterior of $\bm \theta$
\begin{eqnarray}
q\left( \bm \theta_{(\tau)} \right) & \equiv & q\left( \bm \lambda_{(\tau)}, \bm f_{(\tau)}, \bm z_{(\tau)}, \bm \sigma_{(\tau)} \right) \\
& = & q\left( \bm \lambda_{(\tau)} \right)q\left(\bm f_{(\tau)} \right)q\left(\bm z_{(\tau)} \right)q\left( \bm \sigma_{(\tau)} \right) 
\end{eqnarray}
Under this scheme, it can be shown using calculus of variations that maximization of the ELBO is achieved by calculating for each of the four partitions the expectation (w.r.t. all other three partitions) of the log-joint density $\log p \left( \bm \theta_{(\tau)}, \mathbf{x} \right)$. The need to update each partition of the parameters conditional on all others calls for an iterative scheme (similar to the EM algorithm) known as Coordinate Ascent Variational Inference (CAVI).

\begin{figure}[ht]
 \centering
  \begin{minipage}{.75\linewidth}
\begin{algorithm}[H]\footnotesize
\caption{\textit{Variational Bayes Quantile Factor Analysis (VBQFA)}}\label{algorithm:CAVI}
\begin{algorithmic}[0]
\State Initialize $\bm \lambda_{(\tau)}, \bm f_{(\tau)}, \bm z_{(\tau)}, \bm \sigma_{(\tau)}$, set $ELBO^{(0)} = -1000$, and $c = 1e-6$ the tolerance.
    \State $r = 1$
	\While{$ \vert ELBO^{(r)} - ELBO^{(r-1)} \vert > c $}
	    \For{$\tau \in (0,1)$}
		\State Update $q\left( \bm \theta_{(\tau)} \right)$ by sequentially calculating
		\State 1: $q\left(\bm f_{(\tau)} \right) \propto \exp\left \lbrace \mathbb{E}_{q(\bm \lambda_{(\tau)}, \bm z_{(\tau)},\bm \sigma_{(\tau)})}\left[\log p\left( \bm \theta_{(\tau)} , \mathbf{x} \right) \right] \right\rbrace$
		\State 2: $q\left(\bm \lambda_{(\tau)} \right) \propto \exp\left \lbrace \mathbb{E}_{q(\bm f_{(\tau)}, \bm z_{(\tau)},\bm \sigma_{(\tau)})}\left[\log p\left( \bm \theta_{(\tau)} , \mathbf{x} \right) \right] \right\rbrace$
		\State 3: $q\left(\bm z_{(\tau)} \right) \propto \exp\left \lbrace \mathbb{E}_{q(\bm f_{(\tau)}, \bm \lambda_{(\tau)},\bm \sigma_{(\tau)})}\left[\log p\left( \bm \theta_{(\tau)} , \mathbf{x} \right) \right] \right\rbrace$
		\State 4: $q\left(\bm \sigma_{(\tau)} \right) \propto \exp\left \lbrace \mathbb{E}_{q(\bm f_{(\tau)}, \bm \lambda_{(\tau)},\bm z_{(\tau)})}\left[\log p\left( \bm \theta_{(\tau)}, \mathbf{x} \right) \right] \right\rbrace$
	    \EndFor
        \State $r = r + 1$
	\EndWhile
\end{algorithmic}
\end{algorithm}
\end{minipage}
\end{figure}

The quantities in steps 1-4 of the algorithm above may not necessarily be simple and tractable. However, notice that these conditional expectations mean that we evaluate the $p \left( \bm \theta_{(\tau)}, \mathbf{x} \right)$ conditional on fixing the values of three out of the four blocks of parameters at a time. Therefore, this algorithm ends up depending on quantities that are similar to the conditional posteriors derived in Gibbs sampler algorithms, which are typically simple densities. For example, in step 1, by fixing the values of $\bm \lambda_{(\tau)}, \bm z_{(\tau)},\bm \sigma_{(\tau)}$ to their expectations reduces the log joint density to an expression involving a quadratic term in the parameter of interest, $\bm f_{(\tau)}$. By further taking the exponential of this term, the variational density $q\left(\bm f_{(\tau)} \right)$ becomes proportional to the normal distribution, with posterior mean and variance identical to the formulas we would derive for a Gibbs sampler algorithm. In step two, we fix $\bm f_{(\tau)}$ to its posterior mean and proceed to update $\bm \lambda_{(\tau)}$. Therefore, even if some derivations seem cumbersome, the final algorithm outlined above has a very simple structure that should seem natural to end users who are familiar with the Gibbs sampler. The online supplement provides detailed derivations.

In general, variational Bayes is expected to be much faster than the Gibbs sampler and comparable MCMC techniques \citep{Gunawanetal2021}. The algorithm outlined above will typically converge after a handful of iterations (see also \autoref{ELBO1} and associated discussion), whereas MCMC methods rely on iteratively sampling thousands or even millions of times. Our deterministic CAVI-based algorithm's convergence is excellent because all variational distributions involved are conjugate. However, a well-known drawback of deterministic optimization algorithms is that they do not scale well to massive datasets, such as in the analysis of text data found in natural language processing. In these situations, stochastic variational inference (SVI, \citealp{Hoffmanetal2013}) is a much faster option. Compared to CAVI, SVI randomly updates the ELBO to find the closest fit to the real posterior distribution. However, for the macroeconomic and financial time series datasets we are interested in, our variational Bayes algorithm converges quickly, making the derivation of an SVI variant unnecessary. As we describe in the following section, another advantage of our CAVI algorithm is that it can easily be adapted to more flexible model specifications. SVI algorithms, on the other hand, may converge slowly for complex models and require several passes over the data \citep{AlivertiRusso2022}.

\subsection{\textit{Extensions of the model}}
Compared to nonparametric methods, a major benefit of using a probabilistic approach to quantile factor analysis is the ability to incorporate numerous flexible modeling features. 
For example, in traditional factor models, an active literature suggests Bayesian priors that shrink the loadings and help select the number of factors. \cite{West2003} suggests a continuous spike and slab prior, which  \cite{Carvalhoetal2008} extend to accommodate very sparse genome data. Similarly, \cite{KnowlesGhahramani2011} and \cite{RockovaGeorge2016} extend this idea to a nonparametric Bayesian setting, using the spike and slab prior to find zero restrictions on factor loadings. Analogously, \cite{BhattacharyaDunson2011} proposes priors that both shrink loadings and select the unknown number of factors. In the simulation study we explore several of these classes of priors within the context of quantile factor analysis using variational Bayes inference; see \autoref{sec:MC_priors} for more details.

Another feasible extension incorporates time-varying loadings and/or idiosyncratic variances. When performing quantile factor analysis using macroeconomic data, it is not hard to imagine situations where the distribution of the data has shifted to such an extend, that the loadings at different quantile levels are subject to breaks. In the ``mean'' factor model, breaks in the loadings are conveniently represented using time-varying parameters \citep{StockWatson2002a}. Modified this way, the quantile factor model is given by
\begin{eqnarray}
x_{it} & = & \bm \lambda_{it,(\tau)}^{\prime} \bm f_{t,(\tau)} + u_{it,(\tau)}, \\
\bm \lambda_{it,(\tau)} & = & \bm \lambda_{i,t-1,(\tau)} + \bm \eta_{it},
\end{eqnarray} 
where $\bm \eta_{it} \sim N_{r} \left( \bm 0,\bm V \right)$ and $\bm V$ is a diagonal covariance matrix. Similarly, another important feature of time series data is stochastic volatility in $\sigma_{i,(\tau)}$. \cite{Gerlachetal2011} show how it can be integrated into a quantile regression model, such that the variance of each quantile of the data distribution varies over time. From a computational perspective, extensions with time-varying parameters and stochastic volatility are feasible using variational Bayes inference; see \cite{KoopKorobilis2023} for an application to high-dimensional regression models.

Time-varying parameters and stochastic volatility are empirically plausible extensions that accommodate features of modern macroeconomic and financial time series data; however, flexibility does not always equate to better fit. In particular, when the focus lies on forecasting and out-of-sample projections, parsimonious models might be hard to beat. Therefore, in the remainder of this paper, we try to streamline our empirical evidence by focusing only on the linear quantile factor analysis model. In other words, we show that the suggested linear quantile factor estimator can be numerically better than linear PCA and the quantile factor estimator of \citetalias{Chenetal2021}, leaving studies on the benefits of time-varying parameters in quantile factor analysis to future research.

\section{Simulation studies}

In this section, we explore the properties of the VBQFA estimator in a series of synthetic data experiments. In the first exercise we focus on comparing the performance of our probabilistic estimator to the loss-based estimator of \citetalias{Chenetal2021} in recovering the true factors for a variety of error distributions. The next exercise varies the signal-to-noise ratio and explores how performance is affected by signal strength. As an additional benchmark, we estimate our proposed quantile factor model with MCMC to illustrate how signal strength impacts the accuracy of our approximate variational estimator. The third exercise focuses on methods for estimating the number of factors. In this exercise we demonstrate how the ELBO can be exploited for this purpose and provide additional results on information criteria. Finally, a large set of additional simulations is included in the online supplement: an exercise exploring the performance of VB for various prior distributions compared to MCMC, a simulation exercise in which the residuals are autocorrelated and cross-correlated across series following \citetalias{Chenetal2021}, and a simulation exercise that evaluates the performance of our estimator for a selection of outlier processes, based on \citep{Tsayetal2000}. 

\subsection{\textit{Numerical Precision of the new Estimator}} \label{sec:MC}
\subsubsection*{\textit{Comparison using different error distributions}}
First, we report the results of synthetic data experiments that illustrate the performance of the proposed probabilistic estimator (VBQFA) in recovering factors that are generated from flexible error distributions. In our primary experiments we follow the benchmark setting of \citetalias{Chenetal2021} that imposes the same factor structure for each quantile of the distribution of the synthetic time series $\mathbf{x}_{t}$.\footnote{In the online supplement, we report Monte Carlo results under an alternative scheme considered by \cite{Chenetal2021}, namely the case of dependent idiosyncratic normal and Student t errors. In this scenario, our algorithm and the algorithm of \citetalias{Chenetal2021} provide numerically identical performance.} In order to perform a more broad exploration compared to \citetalias{Chenetal2021}, we consider a series of six flexible distributions that feature fat-tails, skewness, and bimodalities. The generative model for the synthetic data takes the following form:
\begin{eqnarray}
x_{it} &  =  & \sum_{j=1}^{3} \lambda_{ji} f_{jt} + u_{it}, \label{DGP1} \\
f_{jt} &  =  &  0.8 f_{jt-1} + \epsilon_{jt}, j=1,2,3,  \label{DGP2}
\end{eqnarray}
where $\lambda_{ji}$ and $\epsilon_{jt}$ are independent draws from a $N(0,1)$ distribution, and $u_{it}$ is generated from the following distributions
\begin{enumerate}[noitemsep]
\item[M1] Heavy-tailed: $t_{3}(0,1)$
\item[M2] Kurtotic: $2/3 N(0,1^2) + 1/3 N(0,(1/10)^2)$
\item[M3] Outlier : $1/10 N(0,1^2) + 9/10 N(0,(1/10)^2)$
\item[M4] Bimodal : $1/2 N(-1,(2/3)^2) + 1/2 N(1,(2/3)^2)$
\item[M5] Bimodal, separate modes: $1/2 N(-3/2,(1/2)^2) + 1/2 N(3/2,(1/2)^2)$
\item[M6] Skewed bimodal: $3/4 N(-43/100,1^2) + 1/4 N(107/100,(1/3)^2)$
\end{enumerate}
Cases M1-M6 are considered in several quantile regression papers in the statistics literature \citep[see for example the simulation design of][]{Limetal2020}. For reference, the online supplement provides visual comparison of the shapes of the flexible distributions outlined above. We generate 1000 datasets for all combinations of  $T={50,100,200}$ and $n={50,100,200}$. As equation \eqref{DGP1} implies, the true number of factors is fixed to $r=3$.

\begin{figure}
\centering
\subcaptionbox{M1\label{dotsM1}}
{\includegraphics[width=0.32\textwidth, trim={0cm 0cm 0cm 0cm},clip]{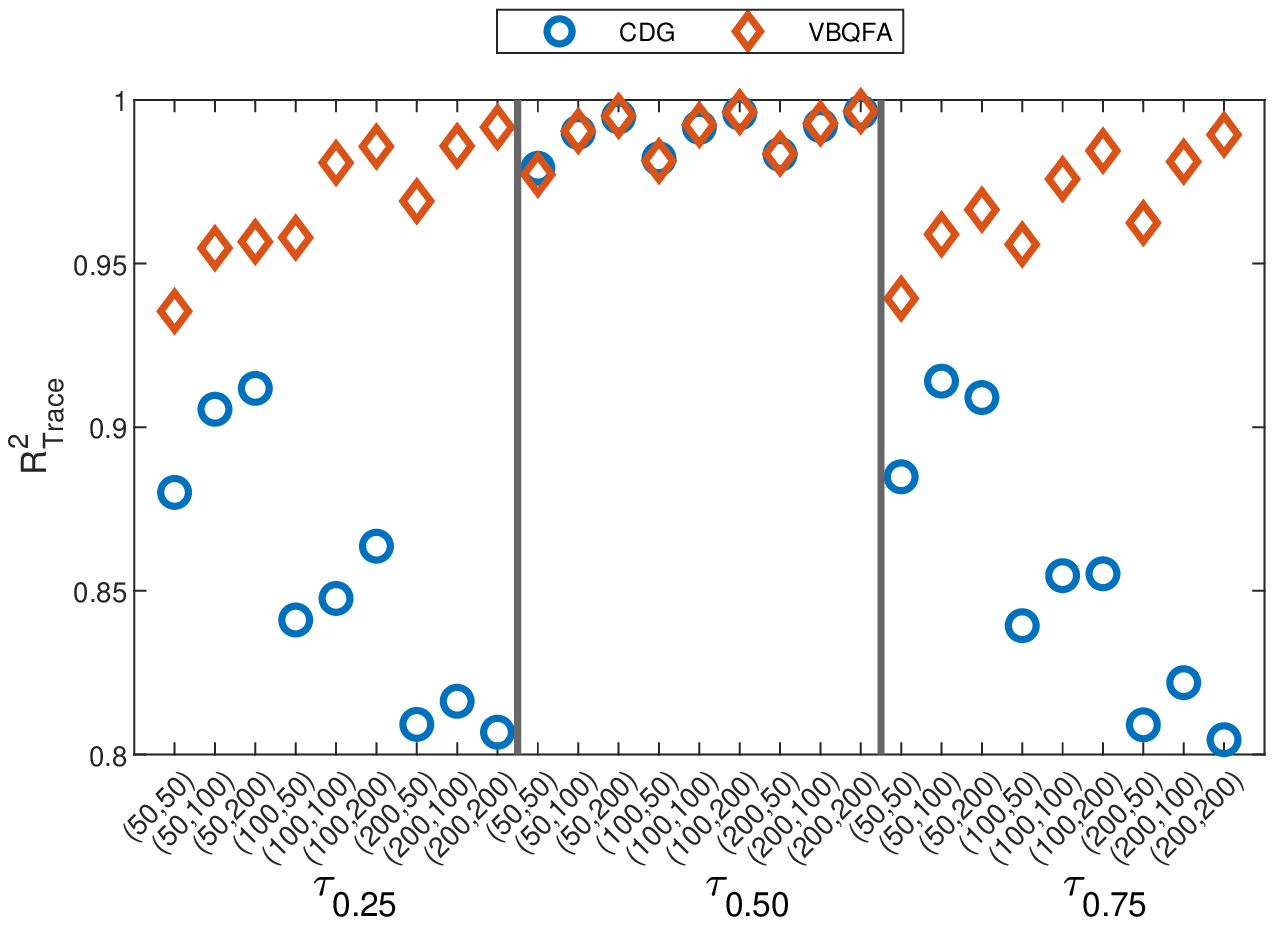}}
\subcaptionbox{M2\label{dotsM2}}
{\includegraphics[width=0.32\textwidth, trim={0cm 0cm 0cm 0cm},clip]{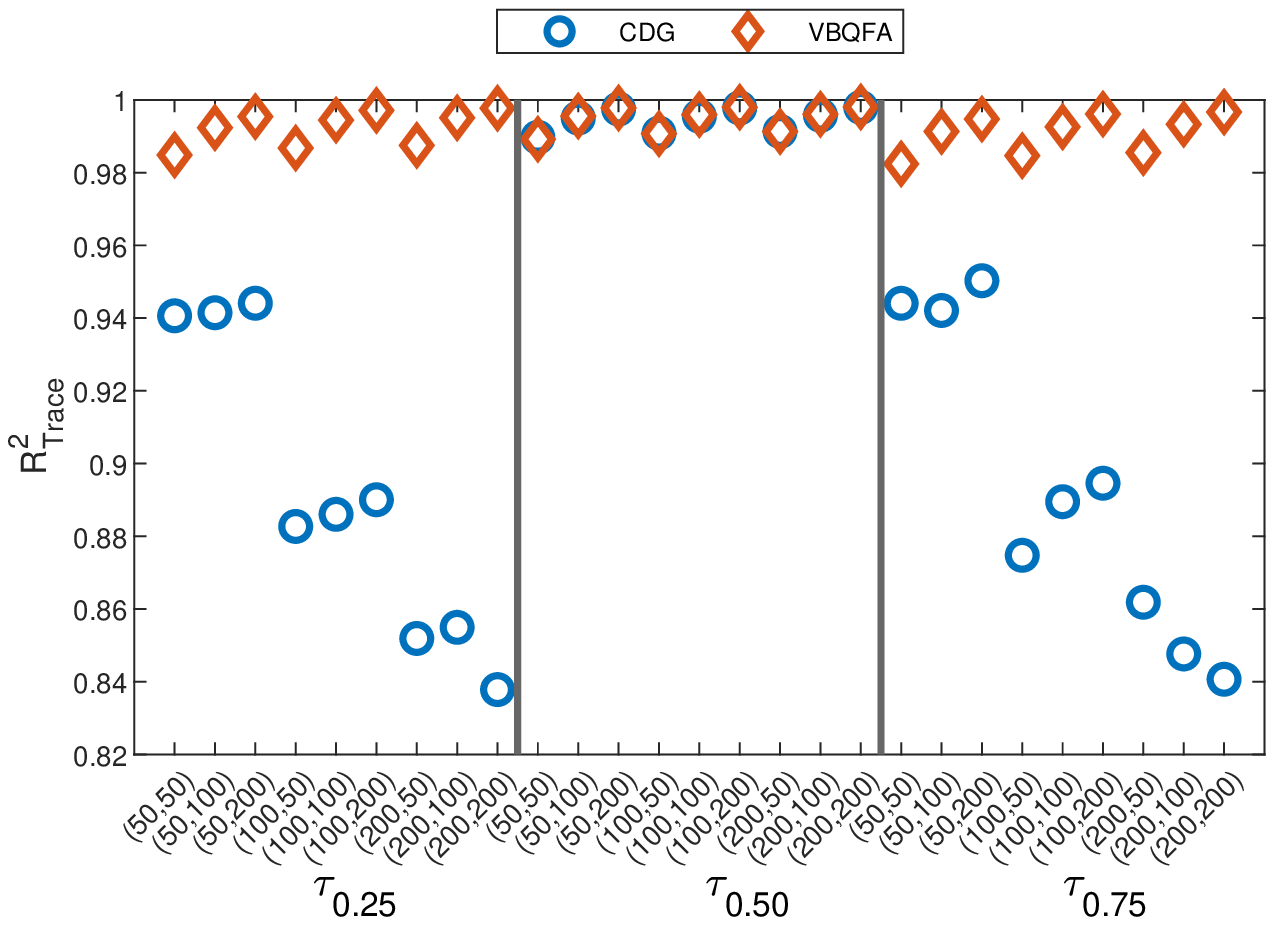}}
\subcaptionbox{M3\label{dotsM3}}
{\includegraphics[width=0.32\textwidth, trim={0cm 0cm 0cm 0cm},clip]{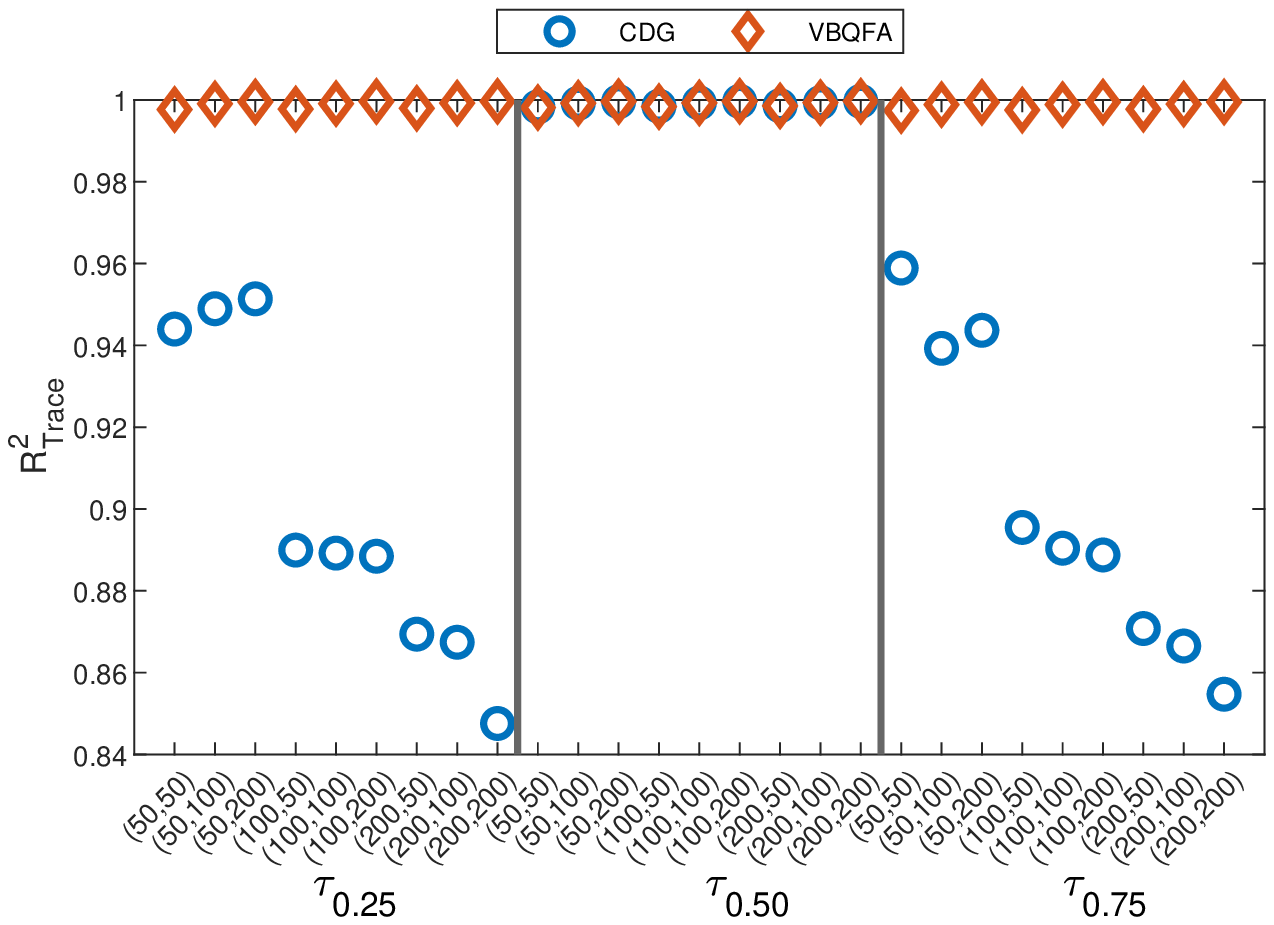}}
\subcaptionbox{M4\label{dotsM4}}
{\includegraphics[width=0.32\textwidth, trim={0cm 0cm 0cm 0cm},clip]{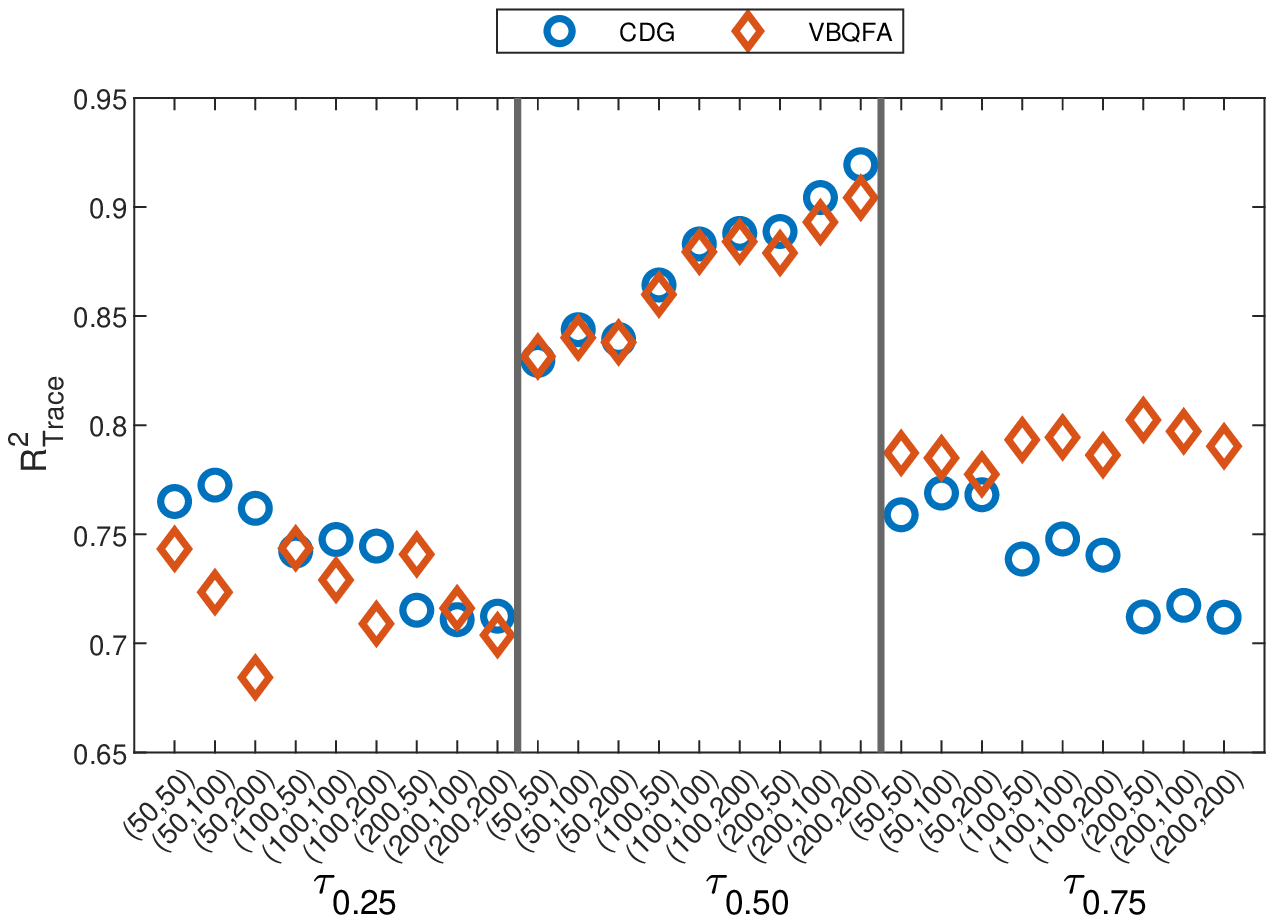}}
\subcaptionbox{M5\label{dotsM5}}
{\includegraphics[width=0.32\textwidth, trim={0cm 0cm 0cm 0cm},clip]{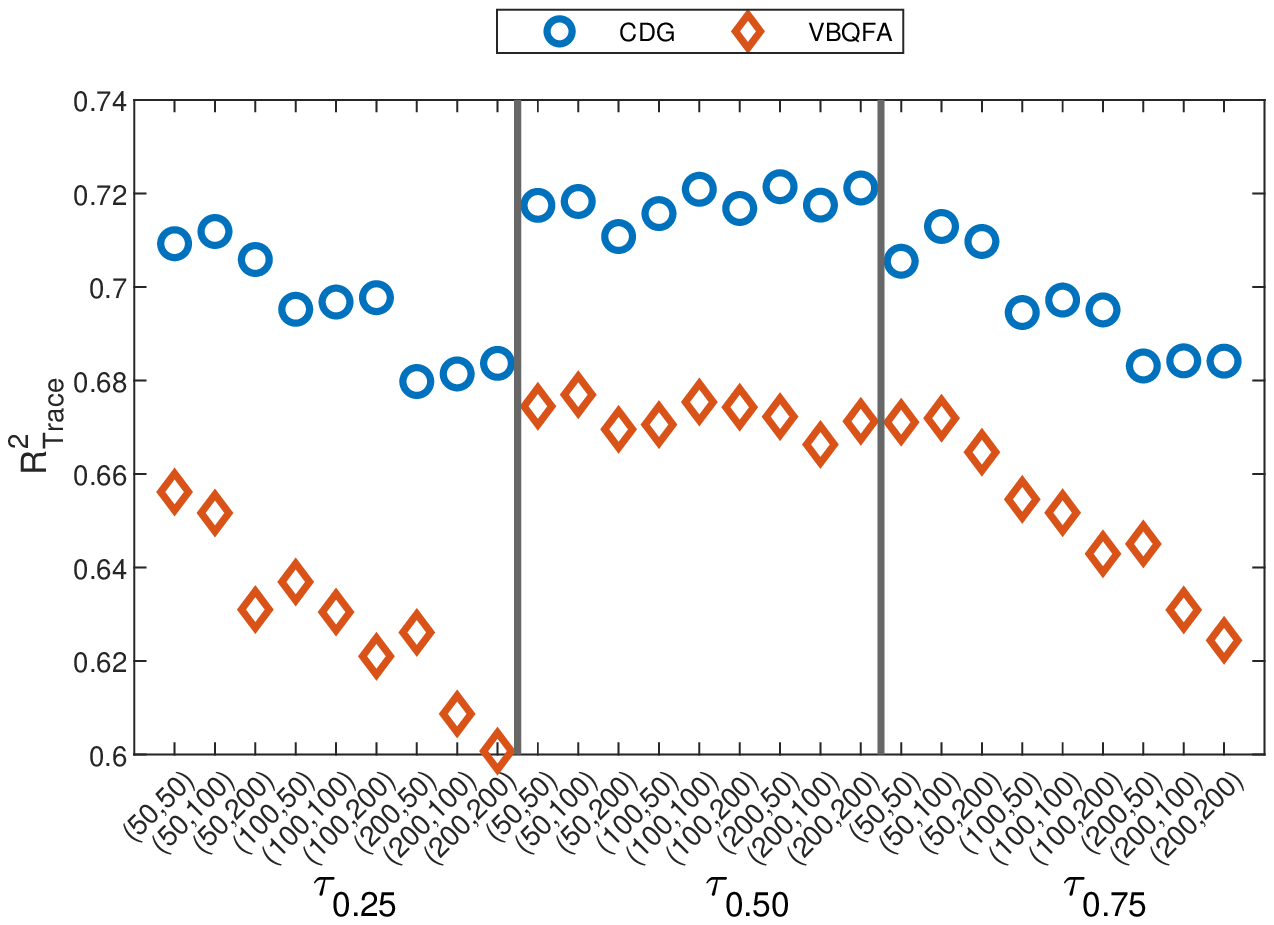}}
\subcaptionbox{M6\label{dotsM6}}
{\includegraphics[width=0.32\textwidth, trim={0cm 0cm 0cm 0cm},clip]{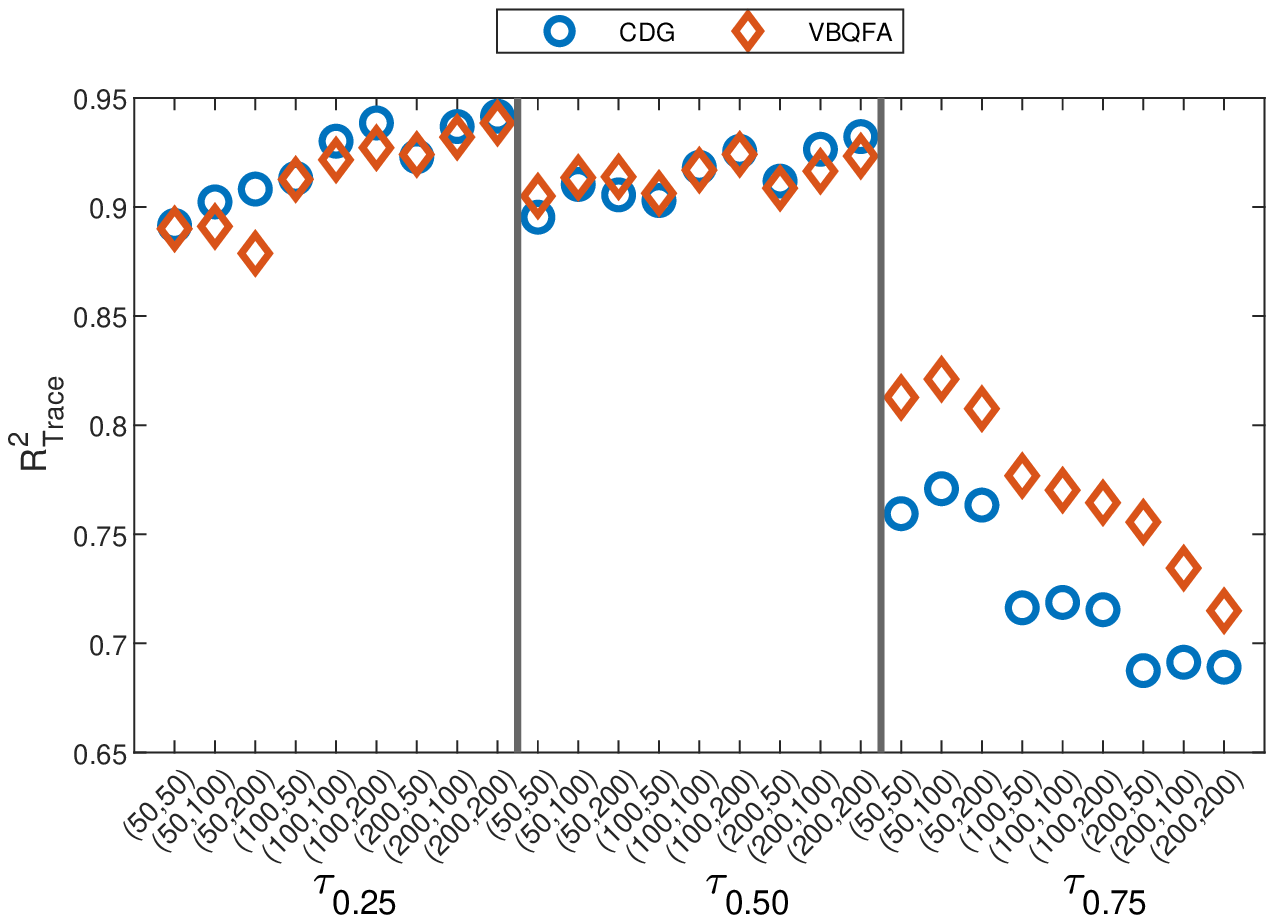}}
\caption{The different panels depict the trace $R^2$ of the multivariate regression of $F$ onto $\hat{F}$ for the \citetalias{Chenetal2021} and VBQFA algorithms for the different data generating processes M1 to M6. The x-axis displays the different pairs of values of $(n,T)$, for each quantile level $\tau_{0.25}$, $\tau_{0.50}$, and $\tau_{0.75}$, respectively. The quantile levels are separated by solid vertical lines. }\label{fig:tracer2}
\end{figure}

Figure \ref{fig:tracer2} provides a visual summary of the results. For each DGP the individual panels show the trace $R^2$ of the multivariate regression of the estimated factors onto the simulated factors for the proposed VBQFA model and the benchmark model developed by \citetalias{Chenetal2021}.\footnote{As explained in \cite{StockWatson2002a}, the trace $R^2$ is given by $R^2_{\hat{\bm f},\bm f} = \hat{\mathbb{E}} tr(\hat{\bm f}'\bm P_f \hat{\bm f})/\hat{\mathbb{E}}tr(\hat{\bm f}'\hat{\bm f})$, where $ \hat{\mathbb{E}}$ is the expectation operator estimated as the average over Monte Carlo iterations, $\hat{\bm f}$ denotes the estimated factors, $\bm f$ denotes the true generated factors, and $\bm P_f = \bm f(\bm f'\bm f)^{-1}\bm f'$. We generalize this measure in order to obtain its value for each quantile level $\tau \in (0,1)$. As with all $R^2$ measures, the trace $R^2$ is bounded by one and higher values signify better statistical fit.} For the heavy-tailed (M1), kurtotic (M2), and outlier distribution (M3) the VBQFA and \citetalias{Chenetal2021} algorithms perform equally well in the median. In the tails, the VBQFA performs much better than \citetalias{Chenetal2021}. Surprisingly, the latter algorithm's performance deteriorates as $n$ increases. For the more complicated distributions in M4 to M6, the picture is mixed. In M4 and M6, VBQFA dominates the right tail; however, for case M5 \citetalias{Chenetal2021} performs consistently better across quantile levels. Overall, except for this bimodal distribution, VBQFA outperforms in accurately recovering the true factors. The online supplement provides detailed tables with the numerical values of the trace $R^{2}$ plotted in \autoref{fig:tracer2}, as well as comparisons using alternative statistical accuracy metrics, and additional robustness exercises.\footnote{Depending on the case, the error distributions simulated in this exercise imply non-zero constants in the tails and median. This fact implies a potential misspecification of the estimation algorithm in \citetalias{Chenetal2021} which does not allow for a constant during estimation. As an additional robustness exercise, we follow the approach to estimation on real data in \citetalias{Chenetal2021} and standardize the simulated data before estimation with their algorithm. The overall conclusions remain the same and lend further support to our proposed approach.}

\subsubsection*{\textit{Comparison using different signal-to-noise ratios}} 
\begin{figure}[H]
\centering
\subcaptionbox{$q=0.25$}
{\includegraphics[width=0.32\textwidth, trim={8cm 0cm 9cm 0cm},clip]{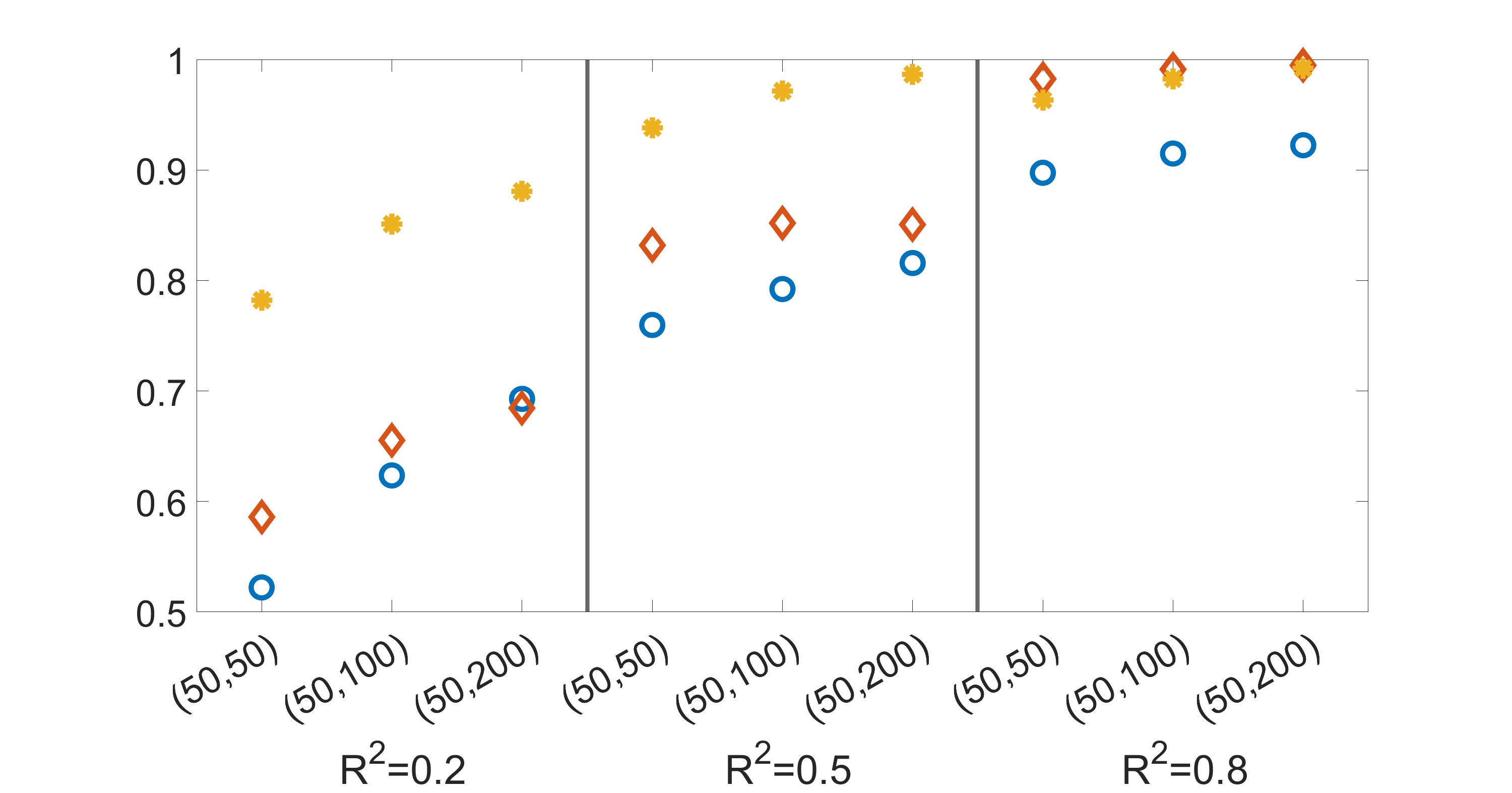}}
\subcaptionbox{$q=0.50$}
{\includegraphics[width=0.32\textwidth, trim={8cm 0cm 9cm 0cm},clip]{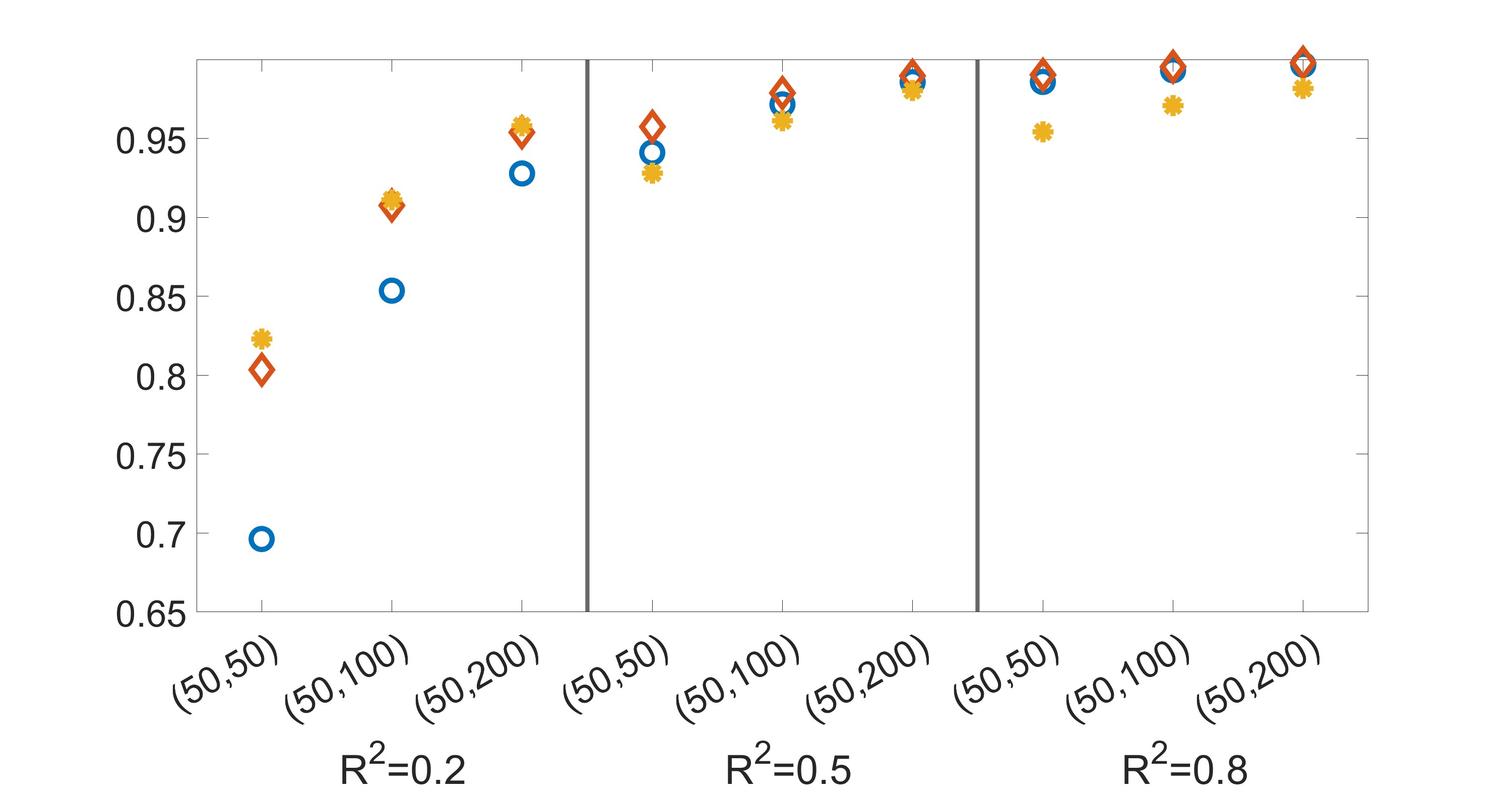}}
\subcaptionbox{$q=0.75$}
{\includegraphics[width=0.32\textwidth, trim={8cm 0cm 9cm 0cm},clip]{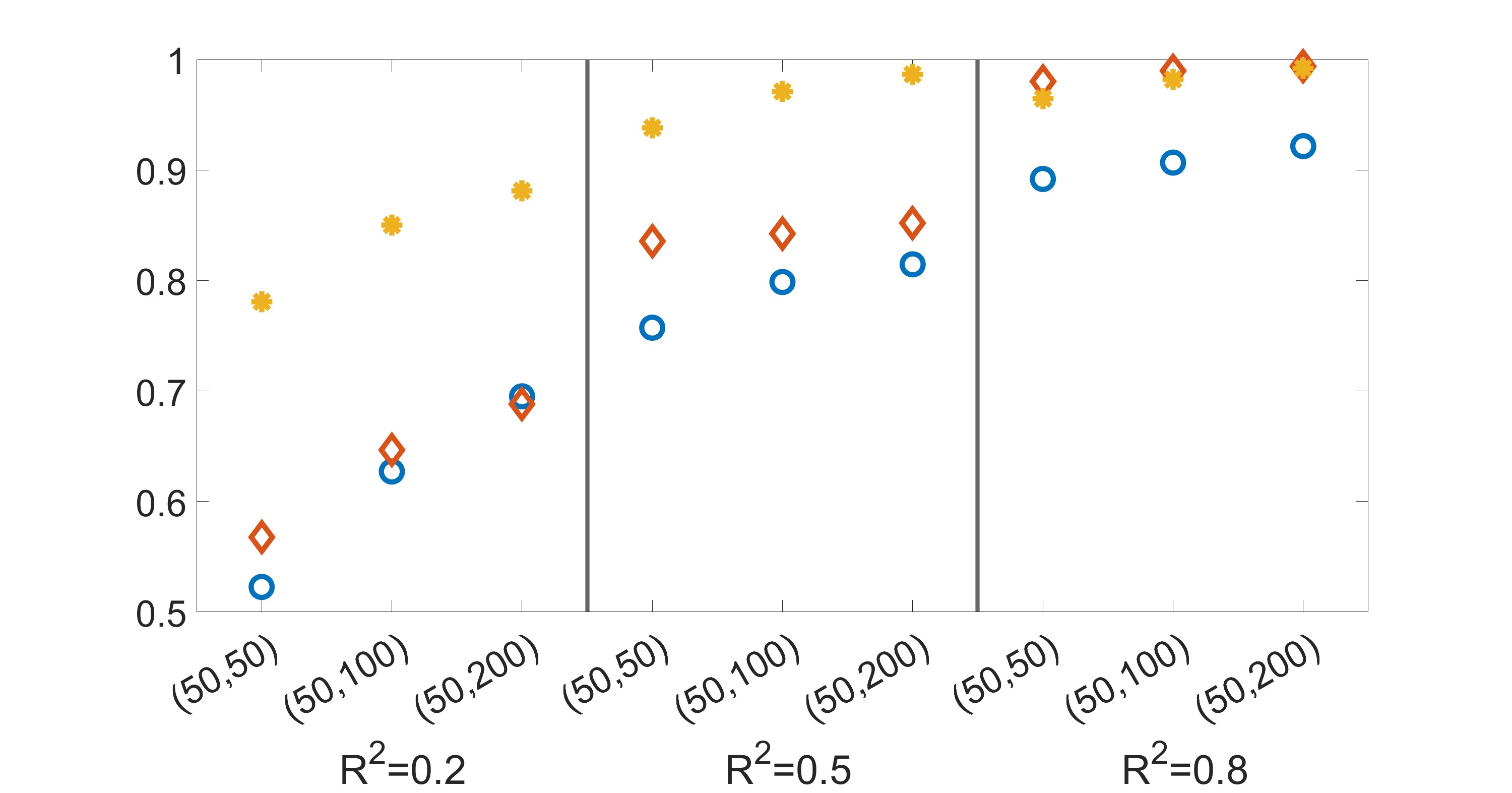}}
\caption{The figure depicts the trace $R^2$ for the three quantile levels $q=[0.25,0.5,0.75]$. The x-axis indicates the simulation size with $(T,N)$. The blue circle indicates the model in \citetalias{Chenetal2021}, the red diamonds the proposed VB estimator and the yellow asterisks the MCMC estimator. The vertical black lines separate the three simulated signal-to-noise ratios $R^2=[0.2,0.5,0.8]$.} \label{SIM:SNR_results}
\end{figure}

In the second experiment, we examine the impact of factor strength on estimation accuracy, especially in the tails. We repeat estimation of model M1,\footnote{Note that the relative performance of the VBQFA and \citetalias{Chenetal2021} might depend on the choice of the DGP.} but now tuning the signal-to-noise ratio of the common factors relative to the idiosyncratic component, in order to achieve three levels of fit (low, medium, high). Specifically, these levels of fit are calibrated such that the common component achieves $R^2 = [0.2,0.5,0.8]$, respectively. \autoref{SIM:SNR_results} presents trace $R^2$ values over 500 Monte Carlo iterations for different combinations of $\tau, T, n$. In addition to the model in \citetalias{Chenetal2021}, we also include results from a Markov chain Monte Carlo (MCMC) algorithm for the probabilistic quantile factor model presented above, which is provided in the online supplement. This comparison allows to evaluate the loss of information when using variational Bayes relative to the more accurate MCMC and how this loss might vary with signal strength.

For the median, VB and MCMC have very similar performance, with VB performing slightly better for higher signal-to-noise ratios. The estimator in \citetalias{Chenetal2021} extracts the true factors less accurately for low signal-to-noise ratios, but performs on par with VB for medium and high signal-to-noise ratios. In the tails, the estimator in \citetalias{Chenetal2021} has the lowest trace R2 in the vast majority of cases. In contrast, VB recovers the true factors less accurately than MCMC for low signal-to-noise ratios, converging to the performance of MCMC as the signal-to-noise ratio increases. While the results might vary across DGPs, they provide some evidence that little accuracy is sacrificed when using VBQFA in empirical applications, especially for higher signal-to-noise ratios. 

\subsection{\textit{Convergence of ELBO and Factor Selection}}

Many applications rely on a fixed number of factors, making factor number selection a secondary issue. For example, in our empirical illustrations, we deliberately extract one uncertainty indicator and one financial condition index, because decision-making often necessitates monitoring a single index that summarizes the information contained in multiple series. Moreover, when interested in forecasting a variable $y$, selecting the number of factors based on in-sample fit criteria is less informative. First, the criteria are computed for factors that are extracted from the data $\bm x$ without direct reference to the target variable $y$. Second, in forecasting applications out-of-sample performance measures (e.g., mean squared forecast errors) are always more appropriate than in-sample measures of fit, also for selecting the number of factors.

However, when factor number selection is an important part of statistical inference, the proposed variational Bayes algorithm can directly yield estimates, as is illustrated in this exercise. Iterating through the steps of \hyperref[algorithm:CAVI]{Algorithm 1} results in the maximization of the evidence lower bound $ELBO$. Given that the $ELBO$ serves as a lower bound to the log-marginal likelihood, it can directly be exploited for model selection. 
To illustrate this, we set up a small experiment and generate artificial data from a three-factor model ($r=3$) with Student-t errors (identical to model M1 of \autoref{sec:MC}), using $T=100$ and $n = 50$. We then proceed to estimate quantile factor models with one, two, three, four, five, and six factors, respectively.

\begin{figure}[H]
\centering
\subcaptionbox{\hfill}
{\includegraphics[width=0.44\textwidth,trim={2cm 1cm 3cm 0cm},clip]{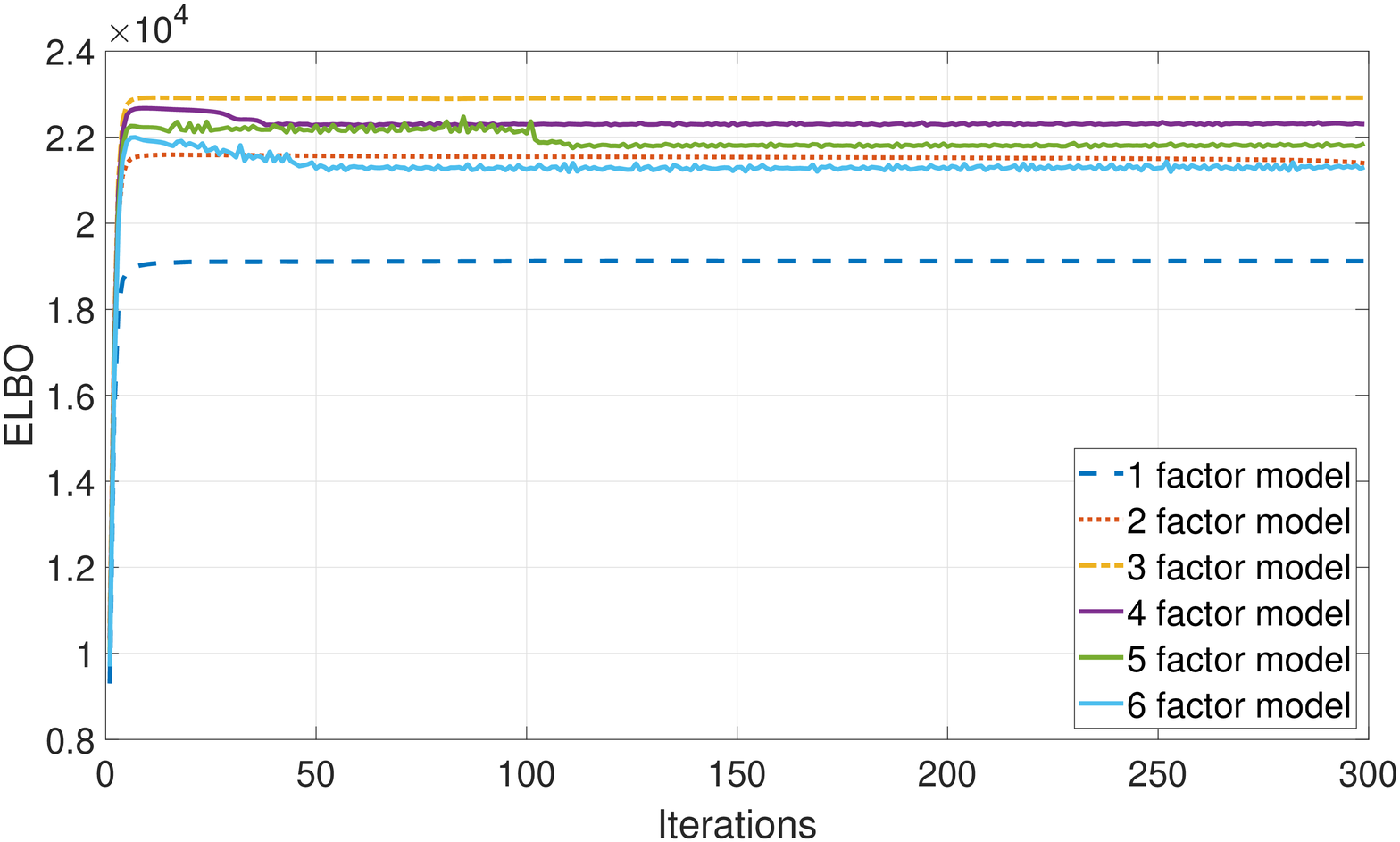}}
\subcaptionbox{\hfill}
{\includegraphics[width=0.44\textwidth,trim={2cm 1cm 3cm 0cm},clip]{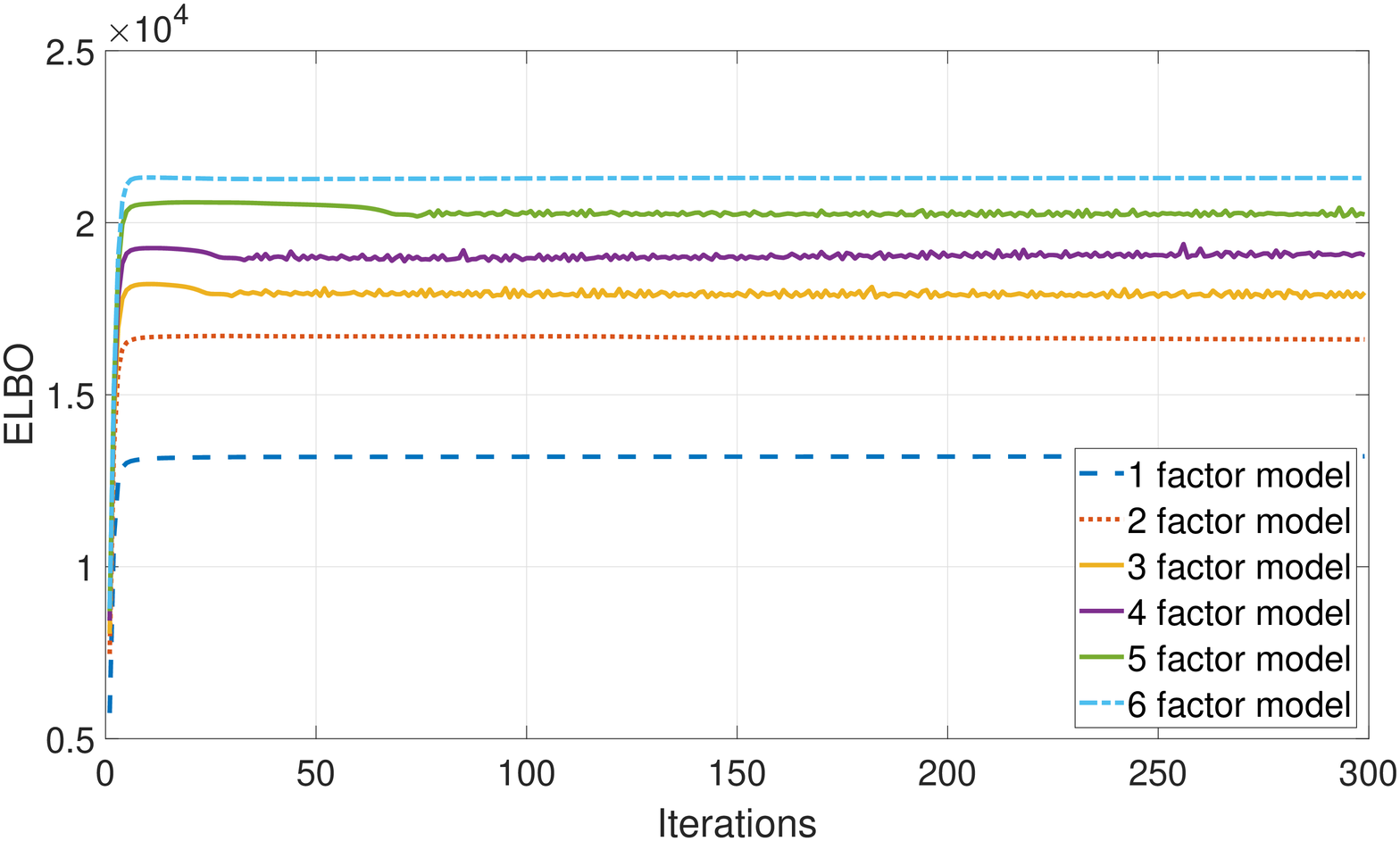}}
\caption{ELBO estimates when the true number of factors in the DGP is $r=3$ (panel (a)) and $r=6$ (panel (b)).} \label{ELBO1}
\end{figure}

Panel (a) of \autoref{ELBO1} shows the evolution of the $ELBO$ over 300 iterations for all six quantile factor model estimates. In all cases, after only a handful of iterations, the $ELBO$ converges towards a fixed value. However, when overfitting the model with four, five, and six factors (solid lines), the $ELBO$ values initially fluctuate before converging to a fixed number. In contrast, when estimating the models with one, two, and three factors, convergence is very fast and smooth. Most importantly, we observe that the estimated model with three factors (yellow dashed-dotted line) achieves the highest ELBO value.

We repeat this experiment, but now simulate a DGP with $r=6$. Panel (b) of \autoref{ELBO1} shows that now the estimated model with six factors is the best, which is consistent with the DGP.

To provide further evidence, section 1.5 in the online supplement presents a full simulation study. The exercise examines how well the $ELBO$ and the twelve information criteria in \cite{BaiNg2002} recover the correct number of factors for different quantile levels, sample sizes, and numbers of simulated factors. Across simulations, the $ELBO$ performs robustly and selects the correct number of factors in between 88\% and 98\% of cases, outperforming the information criteria. Among these, $IC_2$, $PC_2$, and $AIC_3$ perform the best, selecting the correct number of factors in 74\% to 94\% of cases. Large performance differences to the $ELBO$ emerge particularly when the number of simulated factors increases.

\subsection{\textit{Other simulation exercises}} \label{sec:MC_priors}
The online supplement provides further simulation evidence, the results of which we briefly summarize here for the sake of brevity.

First, we compare the performance of VBQFA vs. MCMC using popular shrinkage priors. The previous section defined the sparse Bayesian learning prior of \cite{Tipping2001}, which is our benchmark prior throughout this paper. In a simulation exercise, we also explore the Horseshoe prior in \cite{Carvalhoetal2010}, the Bayesian Lasso in \cite{park2008bayesian}, a student-t version of the Stochastic Search Variable Selection (SSVS) prior in \cite{george1993variable}, and the non-adaptive version of the multiplicative gamma process shrinkage prior in \cite{BhattacharyaDunson2011}. The proposed VB estimator and the corresponding MCMC algorithm generally perform on par for the different prior distributions. In comparison, the model in \citetalias{Chenetal2021} is less accurate than the proposed estimators, especially for the tails. Overall, the results suggest that the proposed algorithm is robust to different shrinkage prior choices and highlight the empirical relevance of regularization for accurately recovering extreme quantiles.

Next, we simulate quantile factor models with serially and cross correlated residuals, replicating the simulation exercise in \citetalias{Chenetal2021}. This study allows to examine how various types of error correlation influence quantile factor estimation. Across different sample sizes and quantile levels, we observe that both the VBQFA and the \citetalias{Chenetal2021} algorithms perform on par in the vast majority of cases. Notably, factor extraction  remains robust to residual correlation for factors that enter the data generating process linearly. This further supports the favorable properties of our estimator identified thus far.

Finally, we provide evidence showcasing the ability of quantile factor analysis to mitigate the impact of outliers. Based on the work of \cite{Tsayetal2000}, we look at multivariate time series that share a common dynamic factor model structure and are interrupted by four types of outliers: i) new, ii) additive, iii) level shifts, and iv) ramp shifts. In the first three cases, VBQFA dominates, but in the case of ramp shifts the \citetalias{Chenetal2021} algorithm performs better. However, note that ramp shifts often lead to trending behavior after the outlier has occurred. In that case, first differencing the series would transform the ramp shift outlier into a level shift \citep{Tsayetal2000}, allowing to leverage the favourable properties of VBQFA in this case as well.

\section{Empirical illustrations}


\subsection{\textit{Constructing quantile diffusion indexes}}
Economists have long used factor models to construct indexes that summarize information from several series, such as leading indicators of economic activity. Since the onset of the global financial crisis in 2007-2008, financial conditions or stress indexes, along with indexes measuring different types of economic uncertainty, have gained popularity. In recent times, the literature has witnessed a proliferation of several additional indexes, encompassing a wide range of concepts such as climate and epidemiological risks, as well as geopolitical risks. Recognizing the complexity of modern economies, such indexes allow for a parsimonious representation of large information sets without the need to rely explicitly on high-dimensional statistical modeling. Indexes can be created by gathering different individual (and potentially partial) measurements of the desired concept, such as financial conditions, and subsequently calculating their weighted average. The weights in this average can be computed either by employing Ordinary Least Squares (OLS) on the first principal component of the series or by the analyst assigning weights based on an economic estimate of the significance of each series.

Constructing the index as a weighted average of the observed time series provides a simple interpretation of economic and financial indexes, but it might not be a statistically optimal way to combine information from different series. Financial and macroeconomic time series have stronger correlations during recessions, bull markets, and other similar episodes. Therefore, it might be the case that linear combinations of different areas of the distribution of the data provide a better measured index, especially when the desire is to construct an index that measures risk, volatility, stress, uncertainty, and similar concepts. Quantile factor analysis allows to do exactly this, that is, extract factors that are linear combinations of individual series at different parts of their distribution (i.e., quantile levels $\tau$).

We test our premise that quantile indexes might be superior to mean (or median) indexes by means of two forecasting exercises that use two popular datasets. The first dataset comprises the Chicago Fed National Financial Conditions Index (NFCI) and its 105 components, spanning various financial concepts measured at weekly frequency. In this case, the NFCI is a weighted average of its components, with weights determined by Chicago Fed staff. The second dataset contains the Economic Policy Uncertainty (EPU) index and its nine categorical uncertainty indexes, as constructed by \cite{Bakeretal2016}, available at monthly frequency. The total EPU and the disaggregated (categorical) indexes are based on the textual analysis of millions of newspaper articles that contain the words ``economic'', ''policy'', and ``uncertainty'' and their derivatives. As both the NFCI and EPU datasets are fairly standard in economic research and we use the original time series provided by the sources without any transformations or other processing (e.g., outlier adjustment), we leave the full description of these two datasets for the online supplement.

\begin{figure}
\centering
\includegraphics[width=.9\textwidth,trim={0cm 0cm 0cm 0cm},clip]{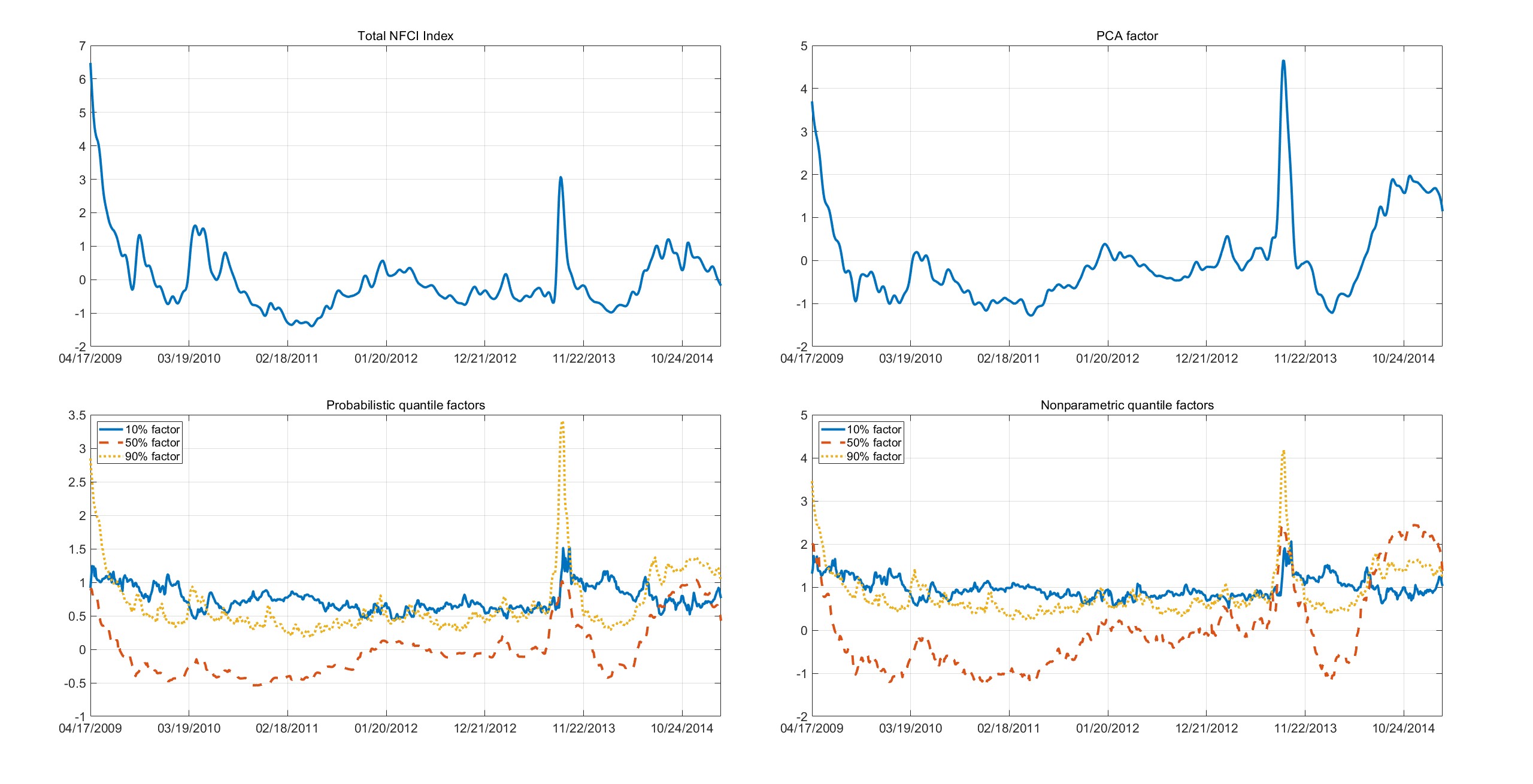}
\caption{Aggregate NFCI vs three different factor estimates using its 105 component series: i) principal component analysis (top right panel); ii) probabilistic quantile factor analysis (bottom left panel); iii) loss-based quantile factor analysis (bottom right panel)} \label{fig:NFCI_factors}
\end{figure}

In both exercises we compare the total index, which is a weighted average, to the first principal component, quantile factors extracted using the \citetalias{Chenetal2021} algorithm, and quantile factors extracted using our VBQFA algorithm. For the quantile factor models we extract indicators at the 10th, 50th and 90th percentiles, that is, we set $\tau={0.1,0.5,0.9}$. We interpret these three percentile levels as representing ``loose'',``medium'' and ``tight'' financial conditions (for the FCI data), or ``low'', ``medium'' and ``high'' uncertainty (for the EPU data). \autoref{fig:NFCI_factors} plots the NFCI series constructed by the Chicago Fed together with the PCA and the two quantile factor estimates extracted from the 105 component series. The data is observed at weekly frequency, spanning from week 3 of 2009 to week 52 of 2023. The total index is visually similar to the principal component, although there are some differences in scale. Compared to these, the quantile factors, whether estimated via \cite{Chenetal2021} or VBQFA, show marked differences depending on the estimated percentile level. While the median factors are almost identical to PCA, the 10th and 90th percentile factors differ dramatically. The 90th percentile factor looks like a smoother variant of the median factor, while the 10th percentile factor has completely different dynamics and peaks compared to all other factors. Apart from small differences in their dynamic evolution, the quantile factors estimated with the probabilistic VBQFA and the loss-based QFA algorithms are visually identical.

\begin{figure}
\centering
\includegraphics[width=.9\textwidth,trim={3cm 1cm 3cm 1cm}]{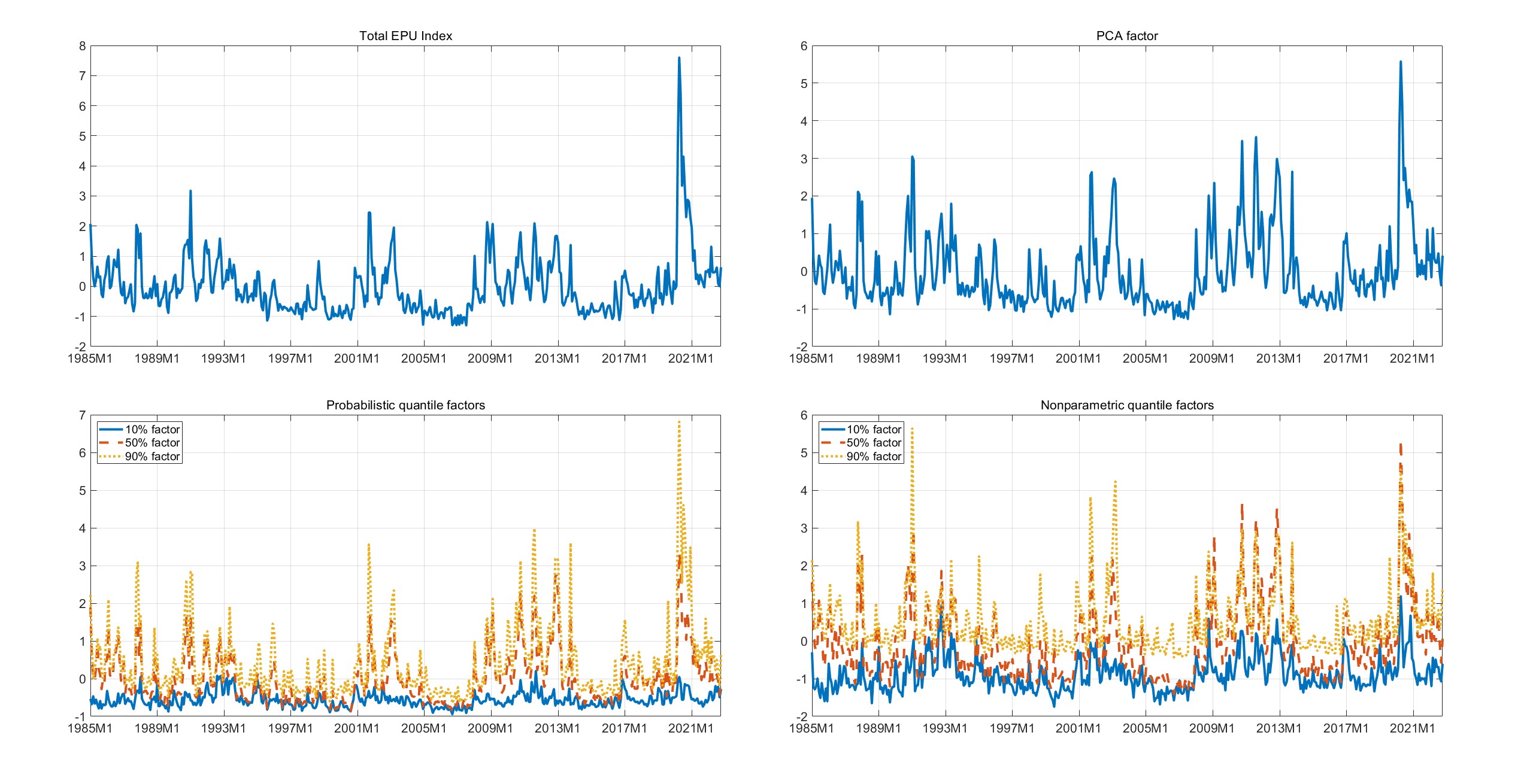}
\caption{Aggregate EPU Index vs factor estimates using categorical EPU series: principal component (top right panel); probabilistic quantile factor analysis (bottom left panel); loss-based quantile factor analysis (bottom right panel)} \label{fig:uncertainty_factors}
\end{figure}

Similarly, \autoref{fig:uncertainty_factors} plots the total EPU index together with the PCA estimate and the two quantile factor estimates extracted from the nine categorical uncertainty series. The sample spans 1985M1 to 2022M10. Here again, the PCA factor is visually identical to the total index, although the scaling of the two series is different, affecting the intensity of different turning points. Also with regards to quantile factors, the median and 90th percentile factors again look comparable, while the peaks on the 10th percentile factor are less pronounced. Finally comparing among different estimation methods, again the series of quantile factors are broadly similar. However, VBQFA and the loss-based QFA algorithm provide quite different estimates of peaks.

Along with the original index, these figures show that for each of the two datasets we can construct seven additional indexes: the PCA, the three nonparametric quantile factors extracted with the \citetalias{Chenetal2021} algorithm, and the three probabilistic quantile factors extracted with VBQFA. We denote these estimates, respectively, as $f^{PCA}$, $f^{CDG}_{0.1}$, $f^{CDG}_{0.5}$, $f^{CDG}_{0.9}$, $f^{VBQFA}_{0.1}$, $f^{VBQFA}_{0.5}$, and $f^{VBQFA}_{0.9}$. In order to check the informational content of each of these indexes, we perform a forecasting exercise where the target variables are measures of economic activity, denoted as $\bm y_{t}$. We use simple OLS to estimate a $p$-lag vector autoregression (VAR) with an intercept on $[y_{t}, index_{t}]$, where $index_{t}$ represents each of the seven factors plus the original index. We then use standard VAR formulas to make predictions $h$ periods into the future \citep {Helmut2005}. We begin the estimation process with the first 50\% of the available sample, forecast $h$ steps ahead, expand the estimation sample by one period, and continue forecasting until we exhaust the entire sample. We then evaluate the forecasts using the second half of the sample (pseudo-out-of-sample evaluation period) minus $h$ observations. Forecast performance is measures separately for each horizon $h$, based on the mean squared forecast error. It is computed as the average of the Euclidean distance between the VAR forecast of $\bm y$ and its realization over the pseudo-out-of-sample evaluation period.


In the first forecasting exercise, we use the NFCI and the seven alternative indexes/factors to forecast the Weekly Economic Index (WEI) maintained by the Federal Reserve Bank of Dallas. The WEI was first constructed during the COVID-19 pandemic to monitor the pandemic's effect on economic conditions in real-time. We collect observations of the WEI for the same period as the FCI data, that is, week 3, 2009 to week 52, 2023. We estimate a VAR with 2 lags where $y$ is the WEI, and we evaluate its forecasts up to $h=26$ weeks (that is, half a year) ahead. \autoref{table:NFCI} presents the mean squared forecast errors (MSFEs) for the WEI for the seven VAR models, estimated using each of the indexes $f^{PCA},f^{CDG}_{0.1},f^{CDG}_{0.5},f^{CDG}_{0.9},f^{VBQFA}_{0.1},f^{VBQFA}_{0.5},f^{VBQFA}_{0.9}$. The MSFEs are compared to the MSFE of a VAR using the WEI and the original NFCI calculated by the Chicago Fed; therefore, values lower (higher) than one indicate a better (worse) performance of an alternative index relative to the NFCI.

\begin{table}[h]
\centering
\resizebox{0.6\textwidth}{!}{
\setlength\extrarowheight{-6pt}
\begin{tabular}{lccccccc} \hline
	&	$f^{PCA}$ 	&	 $f^{CDG}_{0.1}$ 	&	$f^{CDG}_{0.5}$ 	&	$f^{CDG}_{0.9}$ 	&	$f^{VBQFA}_{0.1}$ 	&	 $f^{VBQFA}_{0.5}$ 	&	 $f^{VBQFA}_{0.9}$		\\ \hline\hline
$h=1$	&	0.970	&	1.059	&	1.051	&	0.981	&	1.050	&	1.010	&	0.983		\\
$h=2$	&	0.985	&	1.077	&	1.068	&	0.996	&	1.070	&	1.020	&	0.993		\\
$h=3$	&	1.022	&	1.088	&	1.084	&	1.003	&	1.081	&	1.028	&	0.998		\\
$h=4$	&	1.064	&	1.091	&	1.089	&	1.001	&	1.086	&	1.029	&	0.996		\\
$h=5$	&	1.112	&	1.093	&	1.093	&	1.003	&	1.089	&	1.026	&	0.990		\\
$h=6$	&	1.161	&	1.097	&	1.097	&	1.005	&	1.093	&	1.024	&	0.989		\\
$h=7$	&	1.215	&	1.101	&	1.100	&	1.007	&	1.097	&	1.018	&	0.990		\\
$h=8$	&	1.274	&	1.104	&	1.102	&	1.008	&	1.101	&	1.012	&	0.989		\\
$h=12$	&	1.519	&	1.091	&	1.099	&	1.012	&	1.092	&	0.982	&	0.982		\\
$h=16$	&	1.737	&	1.082	&	1.113	&	1.007	&	1.084	&	0.969	&	0.971		\\
$h=20$	&	2.028	&	1.056	&	1.113	&	1.012	&	1.057	&	0.969	&	0.969		\\
$h=24$	&	2.213	&	1.024	&	1.104	&	0.962	&	1.024	&	0.941	&	0.918		\\
$h=26$	&	2.330	&	1.004	&	1.096	&	0.940	&	1.003	&	0.930	&	0.893		\\ \hline
\end{tabular}
}
\caption{Mean squared forecast errors (MSFEs) of the weekly economic index (WEI), for various weekly forecast horizons $h$. The models used to produce these forecasts are bivariate VARs of WEI and each of the (quantile) factors estimated using PCA, \citetalias{Chenetal2021}, and VBQFA algorithms. MSFEs are given relative to the VAR estimated on WEI and Chicago Fed's original NFCI index.}\label{table:NFCI}
\end{table}

The PCA factor improves upon the forecast performance of the NFCI only slightly for one and two weeks ahead but then its performance deteriorates significantly as the forecast horizon increases. More interesting results emerge for the quantile factor models. For the model in \citetalias{Chenetal2021} performance gains emerge for the 90th percentile factor only for $h=1,2,24,25$, while no gains emerge for the 10th percentile or the median. In contrast, for quantile factors extracted with VBQFA, gains emerge for the median factor for $h>8$ and for all forecast horizons for the 90th percentile factor. Performance improvements range form 0.2 percentage points to 10.7 percentage points. Overall, the 90th percentile factor extracted with VBQFA, which captures particularly tight financial conditions, hence provides important information on real economic activity that is not captured by standard financial conditions indexes, including the NFCI.

In the second forecasting exercise, we use the annualized growth rates of total industrial production and the total consumer price index, as well as the federal funds rate, as the variables of interest. The data are obtained from the Federal Reserve Economic Data (FRED) database and have the mnemonics INDPRO, CPIAUSL, and FEDFUNDS. All series are observed at a monthly frequency from 1985M1 to 2022M10. Other than taking growth rates of output and prices (defined as the first difference of logarithms), we do not apply any other transformation, nor do we adjust for the large outliers observed during the 2020–21 period. In the same way that the previous table showed the FCI forecasts, \autoref{table:OOS_uncertainty} shows the MSFEs for the seven VARs when using each of the different indexes, compared to the VAR's MSFE when using the total EPU index. Given the monthly frequency of the data, we estimate VAR(12) models with least squares and calculate iterated forecasts of up to $h=24$ months ahead.

\vspace*{\fill}
\begin{table}[h]
\centering
\resizebox{1\textwidth}{!}{
\begin{tabular}{lcccccccc|ccccccccc|cccccccc} \hline\hline
	&	 \multicolumn{7}{c}{\textsc{rMSFEs of Industrial Production}}													&	\hspace*{1mm} 	&	\hspace*{1mm} 	&	 \multicolumn{7}{c}{\textsc{rMSFEs of CPI Inflation}}													&	\hspace*{1mm} 	&	\hspace*{1mm} 	&	\multicolumn{7}{c}{\textsc{rMSFEs of Federal Funds Rate}}       													\\ \cmidrule{2-8} \cmidrule{11-17}\cmidrule{20-26}
	&	$f^{PCA}$ 	&	 $f^{CDG}_{0.1}$ 	&	$f^{CDG}_{0.5}$ 	&	$f^{CDG}_{0.9}$ 	&	$f^{VBQFA}_{0.1}$ 	&	 $f^{VBQFA}_{0.5}$ 	&	 $f^{VBQFA}_{0.9}$ 	&	\hspace*{1mm} 	&	\hspace*{1mm} 	&	$f^{PCA}$ 	&	 $f^{CDG}_{0.1}$ 	&	$f^{CDG}_{0.5}$ 	&	$f^{CDG}_{0.9}$ 	&	$f^{VBQFA}_{0.1}$ 	&	 $f^{VBQFA}_{0.5}$ 	&	 $f^{VBQFA}_{0.9}$ 	&	\hspace*{1mm} 	&	\hspace*{1mm} 	&	$f^{PCA}$ 	&	 $f^{CDG}_{0.1}$ 	&	$f^{CDG}_{0.5}$ 	&	$f^{CDG}_{0.9}$ 	&	$f^{VBQFA}_{0.1}$ 	&	 $f^{VBQFA}_{0.5}$ 	&	 $f^{VBQFA}_{0.9}$ 	\\
$h=1$	&	0.965	&	0.996	&	0.981	&	0.978	&	0.993	&	0.988	&	0.953	&	\hspace*{1mm} 	&	\hspace*{1mm} 	&	1.001	&	0.972	&	1.006	&	1.014	&	0.982	&	0.996	&	1.024	&	\hspace*{1mm} 	&	\hspace*{1mm} 	&	0.921	&	0.891	&	0.919	&	0.939	&	0.871	&	0.939	&	0.935	\\
$h=2$	&	0.980	&	0.998	&	1.005	&	0.990	&	0.994	&	1.012	&	0.933	&	\hspace*{1mm} 	&	\hspace*{1mm} 	&	1.000	&	0.971	&	0.989	&	1.025	&	0.965	&	0.990	&	1.033	&	\hspace*{1mm} 	&	\hspace*{1mm} 	&	0.925	&	0.874	&	0.929	&	0.934	&	0.862	&	0.952	&	0.971	\\
$h=3$	&	0.961	&	0.971	&	0.970	&	0.968	&	0.960	&	0.974	&	0.952	&	\hspace*{1mm} 	&	\hspace*{1mm} 	&	1.012	&	0.994	&	1.008	&	1.053	&	0.992	&	1.004	&	1.037	&	\hspace*{1mm} 	&	\hspace*{1mm} 	&	0.937	&	0.883	&	0.946	&	0.936	&	0.868	&	0.977	&	1.001	\\
$h=4$	&	0.975	&	1.008	&	0.992	&	0.965	&	1.006	&	0.982	&	0.958	&	\hspace*{1mm} 	&	\hspace*{1mm} 	&	1.016	&	1.005	&	1.009	&	1.061	&	1.022	&	1.002	&	1.035	&	\hspace*{1mm} 	&	\hspace*{1mm} 	&	0.940	&	0.874	&	0.949	&	0.925	&	0.857	&	0.982	&	1.004	\\
$h=5$	&	1.012	&	1.038	&	1.028	&	1.000	&	1.052	&	1.017	&	0.975	&	\hspace*{1mm} 	&	\hspace*{1mm} 	&	1.016	&	1.002	&	1.009	&	1.048	&	1.019	&	0.994	&	1.030	&	\hspace*{1mm} 	&	\hspace*{1mm} 	&	0.947	&	0.882	&	0.958	&	0.921	&	0.863	&	0.984	&	1.006	\\
$h=6$	&	0.999	&	1.024	&	1.006	&	0.994	&	1.028	&	1.008	&	0.976	&	\hspace*{1mm} 	&	\hspace*{1mm} 	&	1.016	&	1.014	&	1.009	&	1.059	&	1.028	&	0.993	&	1.031	&	\hspace*{1mm} 	&	\hspace*{1mm} 	&	0.962	&	0.899	&	0.975	&	0.926	&	0.882	&	0.999	&	1.017	\\
$h=12$	&	1.008	&	1.026	&	1.019	&	0.999	&	1.021	&	1.006	&	0.992	&	\hspace*{1mm} 	&	\hspace*{1mm} 	&	1.003	&	1.021	&	1.009	&	1.000	&	1.028	&	1.000	&	0.982	&	\hspace*{1mm} 	&	\hspace*{1mm} 	&	1.002	&	0.979	&	1.013	&	0.950	&	0.925	&	1.013	&	1.029	\\
$h=24$	&	1.013	&	1.004	&	1.019	&	1.006	&	1.011	&	1.014	&	1.013	&	\hspace*{1mm} 	&	\hspace*{1mm} 	&	1.014	&	0.999	&	1.021	&	1.012	&	0.987	&	1.010	&	1.010	&	\hspace*{1mm} 	&	\hspace*{1mm} 	&	1.010	&	1.078	&	1.036	&	0.958	&	0.983	&	1.011	&	1.017	\\[1ex]\hline\hline

\end{tabular}
}
\caption{Mean squared forecast errors (MSFEs) of IP, inflation, and FFR, for various monthly forecast horizons $h$. The models used to produce these forecasts are VARs on IP, inflation and FFR and each of the (quantile) factors estimated using PCA, \citetalias{Chenetal2021}, and VBQFA algorithms. MSFEs are given relative to the VAR estimated on the three macro variables and the original EPU index.} \label{table:OOS_uncertainty}
\end{table}

The table reveals some interesting patterns that confirm our premise that indexes constructed from quantile factors can convey more information than mean or median factors. For industrial production, all uncertainty indexes provide marginal predictability in the short-run. The ``mean'' factor ($f^{PCA}$) achieves gains up to $4\%$ versus the aggregate EPU. Looking at the quantile factors, it is not surprising that these forecast gains are only replicated by their $90\%$ factors, while the $10\%$ and $50\%$ factors are on par with total EPU. This observation implies that quantile factors capture periods of high uncertainty (typically associated with recessions) much better than an average index, translating to higher forecast performance gains. It is noticeable that the highest gains, up to almost $7\%$, are achieved by the VBQFA $90\%$ factor. Moving to CPI, the PCA estimate as well as the vast majority of the uncertainty indexes give forecasting results equal to, or worse than, the VAR using the total EPU index. There are only some small gains when using the $10\%$ quantile factors, namely $f^{CDG}_{0.1}$ and $f^{VBQFA}_{0.1}$. Interestingly, the quantile factors seem to affect the interest rate - which is the conventional monetary policy tool - the most. In particular, $f^{PCA}$ as well as the median and $90\%$ factors (both VBQFA and CDG) seem to improve over total EPU by a large margin. However, the largest gains are observed for the $10\%$ quantile factors. This result suggests that, on average, low uncertainty events affect the Fed's decisions more than high uncertainty events. Again, the largest forecasting gains are consistently observed for our VBQFA quantile factor, despite the fact that the \citetalias{Chenetal2021} quantile factor also performs very well.

\subsection{\textit{Extracting quantile factors from large datasets}}


To showcase the VBQFA and its favourable properties in a big data setting, we apply the model to a large data set of daily asset prices. This exercise is different from the previous ones only in that our aim is not to construct a single interpretable index from a set of similar variables. Instead we extract various quantile factors from all available series, with no restrictions on the loadings. The data is directly obtained from \cite{andreou2013should} and spans about 1000 financial time series across five asset classes: commodities, corporate risk, equity, foreign exchange rates, and government securities. The long sample spans the period from 1st January 1986 to 31st December 2008 (4584 trading days) containing 64 time series. The short sample contains 991 daily series from 1st January 1999 to 31st December 2008 (1777 trading days). The series are transformed and normalized following \cite{andreou2013should}.
 
In the original exercise, \cite{andreou2013should} extract factors from the full daily financial data sets as well as individually from the five asset classes. These daily factors are then included individually in mixed-data sampling (MIDAS) regressions to forecast quarterly U.S. real GDP growth. The final forecast is obtained as a weighted average across the factor-MIDAS regressions. 

Similarly, in this exercise three quantile factors are extracted for $q=[0.1,0.5,0.9]$ across all series contained in the long and short sample respectively. These resulting daily factors are then included in MIDAS regressions to forecast quarterly U.S. real GDP growth, where the May 2009 vintage is used. The MIDAS regressions are based on the exponential Almond lag and include a constant, 1 lag of quarterly U.S. real GDP growth, and 2 lags of the MIDAS lag polynomial. 

Following \cite{andreou2013should}, the estimation period for the long and short sample are 1986Q1 to 2000Q4 and 1999Q1 to 2005Q4, whereas the forecasting periods are 2006Q1+h to 2008Q4-h. For the short sample, only one quarter ahead forecasts are considered, while for the long sample one quarter ahead and four quarter ahead forecasts are computed.

\begin{table}[H]
\centering
\resizebox{.7\textwidth}{!}{
\setlength\extrarowheight{-4pt}
\begin{tabular}{lcccc|cccc} \hline\hline
	&	 \multicolumn{3}{c}{\textsc{Simple Average}}	& \hspace*{1mm} 												& \hspace*{1mm}  & \multicolumn{3}{c}{\textsc{Weighted Average}} \\ 
	\hspace*{1mm} & $short$ & $long \text{ } 1q$ & $long \text{ } 4q$ & \hspace*{1mm} & \hspace*{1mm} & $short$ & $long \text{ } 1q$ & $long \text{ } 4q$ \\ \cmidrule{2-4} \cmidrule{7-9} 
\hspace{1cm}$f^{CDG}_{0.1}$	&	1.007	&	0.967	&	0.997	&	\hspace*{1mm}	&	\hspace*{1mm}	&	1.010	&	0.964	&	0.990	\\
\hspace{1cm}$f^{CDG}_{0.5}$	&	0.999	&	0.996	&	1.000	&	\hspace*{1mm}	&	\hspace*{1mm}	&	1.001	&	0.993	&	0.997	\\
\hspace{1cm}$f^{CDG}_{0.9}$	&	0.888	&	0.935	&	0.993	&	\hspace*{1mm}	&	\hspace*{1mm}	&	0.918	&	0.933	&	0.987	\\
\hspace{1cm}$f^{VBQFA}_{0.1}$	&	0.894	&	0.987	&	0.996	&	\hspace*{1mm}	&	\hspace*{1mm}	&	0.903	&	0.982	&	0.992	\\
\hspace{1cm}$f^{VBQFA}_{0.5}$	&	0.988	&	0.978	&	0.984	&	\hspace*{1mm}	&	\hspace*{1mm}	&	0.986	&	0.978	&	0.983	\\
\hspace{1cm}$f^{VBQFA}_{0.9}$	&	0.779	&	0.978	&	0.997	&	\hspace*{1mm}	&	\hspace*{1mm}	&	0.808	&	0.975	&	0.993
\\
\hline \hline
\end{tabular}
}
\caption{The table displays the mean squared forecast errors (MSFE) for the VBQFA and CDG factors for quarterly U.S. real GDP growth relative to the MSFE of the PCA factors. The forecasts are constructed with MIDAS regressions for the high-frequency factors. The left panel shows the results for the simple average and the right panel for the performance based model average. $1q$ and $4q$ denote 1 quarter and 4 quarter ahead forecasts, respectively.} \label{table:MIDAS}
\end{table}

Based on this setup, the final forecasts are computed in two different ways: As a starting point, all factors enter the MIDAS regressions individually. Then forecast combinations are constructed using either the simple average or performance based average following \cite{andreou2013should}. In both cases, the average forecast is constructed separately for each quantile level. Benchmark forecasts based on the first three principal component factors and on the model in \citetalias{Chenetal2021} are computed analogously.

\autoref{table:MIDAS} presents the resulting rMSFE of the forecasts based on the VBQFA and \cite{Chenetal2021} quantile factors relative to the principal component based estimates. The left panel shows the results for the simple average and the right panel for the performance based average. For both, the simple and performance based forecast average, the results are overall similar. In case of the VBQFA, performance gains vis-\'{a}-vis the principal component based forecasts emerge especially for the short sample and for $q=0.1$ and $q=0.9$. In comparison, the model in \citetalias{Chenetal2021} gains compared to the simple benchmark only for $q=0.9$. In addition, this improvement is ten percentage points smaller compared to VBQFA. For both long sample forecasts, both models only improve slightly on the principal component benchmark. Information on tail risk that is particularly useful for forecasting U.S. quarterly real GDP growth might hence be contained in the larger cross-section of asset prices in the short sample, but remain concealed in the long sample.



\section{Conclusion}

We adopt probabilistic inference in the quantile factor analysis model. The model allows to capture complex data distributions by modeling each percentile of that distribution using a separate factor analysis model. We penalize the loadings of each quantile factor by adopting a hierarchical shrinkage prior. We propose a computational fast variational Bayes algorithm that we also demonstrate using many synthetic and real data examples, to be numerically accurate. The empirical illustrations showcase the potential benefits of extracting latent indexes using quantile factor analysis in a variety of settings. Using our methodology we extract quantile financial conditions indexes and uncertainty indicators, as well as quantile factors from large financial data sets. In all cases, our quantile estimates outperform traditional averages, PCA, and other benchmarks in forecasting macroeconomic aggregates.

\bibliographystyle{APA}
\addcontentsline{toc}{section}{\refname}
\bibliography{VBQFA}

\begin{appendix}

\section{Variational Bayes inference in the quantile factor model}
The linear quantile factor model using the mixture-of-normals formulation is
\begin{equation}
x_{it}  = \bm \lambda_{i,(\tau)}^{\prime} \bm f_{t,(\tau)} + \kappa_{1,(\tau)} z_{it,(\tau)}  + \kappa_{2,(\tau)} \sqrt{\sigma_{i,(\tau)} z_{it,(\tau)}} v_{it}, \text{ \ \ } v_{it} \sim N(0,1), \text{ \ \ } \bm f_{t,(\tau)}  \sim  N(\bm 0, \bm I) \label{bqr}
\end{equation}
under the following prior distributions
\begin{equation}
\begin{array}{ccccccc}
\lambda_{ij,(\tau)} \vert \alpha_{ij,(\tau)} & \sim & N(0,\alpha_{ij,(\tau)}^{-1}), & &
\alpha_{ij,(\tau)}                           & \sim & G \left(a_0,b_0\right), \\
\sigma_{i,(\tau)}                            & \sim & G^{-1} \left(r_0,s_0 \right), & &
z_{it,(\tau)}                                & \sim & Exp(\sigma_{i,(\tau)}),
\end{array}
\end{equation}
The set of latent parameters is $\bm \theta_{(\tau)} = \left( \left \lbrace \bm \lambda_{i(\tau)}, \bm \alpha_{i,(\tau)},\bm\sigma_{i,(\tau)} \right \rbrace_{i=1}^{n}, \left \lbrace \bm f_{t,(\tau)}, \bm z_{t,(\tau)} \right \rbrace_{t=1}^{T}\right)$. For convenience we rewrite the objective function of the problem, that is, the evidence lower bound
\begin{equation}
ELBO = \mathbb{E}_{q(\bm \theta_{(\tau)} \vert \mathbf{x})} \left( \log p(\mathbf{x} \vert \bm \theta_{(\tau)})\right) + \mathbb{E}_{q(\bm \theta_{(\tau)} \vert \mathbf{x})} \left( \log p(\bm \theta_{(\tau)}) \right) - \mathbb{E}_{q(\bm \theta_{(\tau)} \vert \mathbf{x})} \left( \log q(\bm \theta_{(\tau)} \vert \mathbf{x})    \right), \label{lower_bound}
\end{equation}
for $\mathbb{E}_{q(\bm \theta_{(\tau)} \vert \mathbf{x})} ( \bullet)$ being the expectation function w.r.t. the variational posterior.

Using calculus of variations, it can be shown that the maximization above can be approximated iteratively by partitioning the parameter set into $M$ independent groups; see \cite{Bleietal2017} for details. Consider the decomposition $q(\bm \theta_{(\tau)} \vert \mathbf{x}) = \prod_{j=1}^{M} q(\bm \theta_{j,(\tau)} \vert \mathbf{x})$, then optimization of the lower bound function can be achieved by sequentially iterating over the densities
\begin{equation}
q(\bm \theta_{j,(\tau)} \vert \mathbf{x}) \propto \exp \mathbb{E}_{q(\bm \theta_{(-j),(\tau)} \vert \mathbf{x})} \left( \log p(\bm \theta_{j,(\tau)} \vert \bm \theta_{(-j),(\tau)}, \mathbf{x}) \right),
\end{equation}
where $\bm \theta_{(-j),(\tau)}$ denotes all elements of $\bm \theta_{(\tau)}$, excluding those in the j$^{th}$ group, $j=1,...,M$. In order to decide on these $M$ groups, we follow \cite{Limetal2020} and assume the following factorization of the variational Bayes posterior density
\begin{align}
q(\bm \theta_{(\tau)} \vert \mathbf{x}) = \prod_{i=1}^{n} \left[q(\bm \lambda_{i,(\tau)} \vert \mathbf{x})  q( \sigma_{i,(\tau)} \vert \mathbf{x}) \prod_{j=1}^{r}                 q( \alpha_{ij,(\tau)}\vert \mathbf{x})  \prod_{t=1}^{T}   q( z_{it,(\tau)} \vert \mathbf{x})  \right] \prod_{t=1}^{T}  q(\bm f_{t,(\tau)} \vert \mathbf{x}), \label{mean_field}
\end{align}
which implies partial posterior independence between certain groups of parameters. Under the additional assumption that the parameter prior distributions are cross-sectionally independent but temporally dependent, the joint prior can be written as \begin{equation}
p(\bm \theta_{(\tau)}) = \prod_{i}^{n} p(\bm \lambda_{i,(\tau)} \vert  \alpha_{i,(\tau)}) p(\bm \alpha_{i,(\tau)})  p(\bm z_{i,(\tau)})p(\bm \sigma_{i,(\tau)}) p(\bm f_{(\tau)} ). \label{prior_decomp}
\end{equation}

In order to capitalize on these decompositions we need to insert equations \eqref{mean_field} and \eqref{prior_decomp} into the formula for the variational lower bound $L$ given in equation \eqref{lower_bound}. Due to this being a lengthy formula (but otherwise easy to derive as it depends on expectations of logarithms of standard densities) we do not present it here. Instead, we focus on deriving expressions for the densities $q(\bm \theta_{j,q} \vert \mathbf{x})$, one at a time for each $j=1,...,M$. Notice that the expectations of each parameter w.r.t. to the variational posterior are defined as: \newline
$\mathbb{E}(\bm \lambda_{i,(\tau)}) = \bm \mu^{\lambda}_{i,(\tau)}$, $\mathbb{E} (\bm \lambda_{i,(\tau)}^2) = \left(\bm \mu^{\lambda}_{i,(\tau)}\right)^2 + diag\left(\bm\Sigma_{i,(\tau)}^{\lambda}\right)$, $\mathbb{E}(\bm f_{t,(\tau)}) = \bm \mu^{f}_{t,(\tau)}$, $\mathbb{E} (\bm f_{t,(\tau)}^2) = \left(\bm\mu^{f}_{t,(\tau)}\right)^2 + diag\left(\bm\Sigma_{t,(\tau)}^{f}\right)$, $\bm \alpha_{i,(\tau)}^{-1} = diag\left(\mathbb{E}  \left(\frac{1}{\alpha_{i1,(\tau)}} \right),...,\mathbb{E} \left( \frac{1}{\alpha_{ik,(\tau)}} \right) \right)$, $\mathbb{E} \left( \frac{1}{\alpha_{ij,(\tau)}}\right) = \frac{a^{\alpha}_{(\tau)}}{b^{\alpha}_{i,(\tau)}}$, $ \mathbb{E} \left(\frac{1}{\sigma_{i,(\tau)}} \right) = \frac{r_{(\tau)}}{s_{i,(\tau)}}$, $\mathbb{E} \left(z_{it}\right) = \frac{\sqrt{b_{it,(\tau)}}K_{3/2} \left( \sqrt{a_{it,(\tau)} b_{it,(\tau)}} \right) }{\sqrt{a_{it,(\tau)}}K_{1/2} \left( \sqrt{a_{it,(\tau)} b_{it,(\tau)}} \right) }$, $\mathbb{E} \left(\frac{1}{z_{it}}\right) = \frac{\sqrt{a_{it,(\tau)}}K_{3/2} \left( \sqrt{a_{it,(\tau)} b_{it,(\tau)}} \right) }{\sqrt{b_{it,(\tau)}}K_{1/2} \left( \sqrt{a_{it,(\tau)} b_{it,(\tau)}} \right) } - \frac{1}{b_{it,(\tau)}}$ and $K_{p}(\bullet)$ is the Bessel function of order $p$.
\begin{algorithm}[h]\scriptsize
{
\caption{\textit{Variational Bayes Quantile Factor Analysis (VBQFA)}}\label{algorithm:VBQFA}
\begin{algorithmic}[1]
\State r = 1 \While{$\vert ELBO^{r} - ELBO^{r-1} \vert \neq 0$}
\State 1) Update $\bm \lambda_{i,(\tau)}$ for $i=1,....,n$ from 
\begin{equation}
q(\bm \lambda_{i,(\tau)} \vert \mathbf{x}) = N_{r}(\bm \mu^{\lambda}_{i,(\tau)}, \bm \Sigma^{\lambda}_{i,(\tau)}),
\end{equation}
where $\bm \Sigma^{\lambda}_{i,(\tau)} = \left( \frac{1}{\kappa_{2,(\tau)}^{2}}\sum_{t=1}^{T}  \mathbb{E} \left(\bm f_{t,(\tau)}\right)  \mathbb{E} \left(\bm f_{t,(\tau)}\right)^{\prime} \mathbb{E} \left(\frac{1}{z_{it,(\tau)}} \right) \mathbb{E} \left( \frac{1}{\sigma_{i,(\tau)}} \right) + \bm \alpha_{i,(\tau)}^{-1} \right)^{-1}$ and $\bm \mu^{\lambda}_{i,(\tau)} = \bm  \Sigma^{\lambda}_{i,(\tau)}\left(\frac{1}{\kappa_{2,(\tau)}^{2}} \sum_{t=1}^{T}  \mathbb{E} \left(\bm f_{t,(\tau)}\right) x_{it} \mathbb{E}\left(\frac{1}{z_{it,(\tau)}} \right) \mathbb{E} \left( \frac{1}{\sigma_{i,(\tau)}} \right) \right. - \left. \frac{\kappa_{1,(\tau)}}{\kappa_{2,(\tau)}^{2}} \mathbb{E} \left( \frac{1}{\sigma_{i,(\tau)}} \right) \sum_{t=1}^T \mathbb{E} \left(\bm f_{t,(\tau)}\right)\right).$ 
\State 2) Update $\bm \alpha_{i,(\tau)}$ for $i=1,....,n$ from 
\begin{equation}
q(\bm \alpha_{i,(\tau)} \vert \mathbf{x}) = G^{-1} \left( a^{\alpha}_{(\tau)}, b^{\alpha}_{i,(\tau)}\right),
\end{equation}
where $a^{\alpha}_{(\tau)} = a_0 + \frac{1}{2}$ and $b^{\alpha}_{i,(\tau)} =  b_0 + \frac{1}{2} \mathbb{E} \left(\lambda_{i,(\tau)}^{2}\right) $.
\State 3) Update $z_{it,(\tau)}$ for $i=1,...,n$ and for $t=1,...,T$ from 
\begin{equation}
q(z_{it,(\tau)}  \vert \mathbf{x}) = IG \left(\frac{1}{2},a_{it,(\tau)},b_{it,(\tau)} \right).
\end{equation}
where $a_{it,(\tau)} = \mathbb{E} \left( \frac{1}{\sigma_{i,(\tau)}}\right) \left( 2 + \frac{\kappa_{1,(\tau)}^{2}}{\kappa_{2,(\tau)}^{2}}\right)$ and $b_{it,(\tau)} = \mathbb{E} \left( \frac{1}{\sigma_{i,(\tau)}}\right)  \frac{\left(\mathbf{x}_{it} -  \mathbb{E}  \left( \bm \lambda_{i,(\tau)}\right) E \left( \bm f_{t,(\tau)}\right)  \right)^2 +  \mathbb{E} \left(f_{t,(\tau)} \right) \bm \Sigma^{\lambda}_{i,(\tau)}  \mathbb{E} \left( f_{t,(\tau)}\right)'  }{\kappa_{2,(\tau)}^{2}}$. Here $IG(\bullet)$ is the three parameter Inverse Gaussian distribution.
\State 4) Update $\sigma_{i,(\tau)}$ for $i=1,...,n$ from
\begin{equation}
q(\sigma_{i,(\tau)}  \vert \mathbf{x}) = G^{-1}(r_{(\tau)},s_{i,(\tau)}),
\end{equation}
where $r_{(\tau)} = r_0 + 3T $ and \begin{align*} s_{i,(\tau)} & = s_0 + \sum_{t=1}^{T} \Biggr[ \mathbb{E} \left( \frac{1}{z_{it,(\tau)}}\right) \frac{\left(\left(\mathbf{x}_{it} -  \mathbb{E} \left( \bm \lambda_{i,(\tau)}\right) \mathbb{E}  \left( \bm f_{t,(\tau)}\right)  \right)^2 +  \mathbb{E} \left( \bm f_{t,(\tau)} \right) \bm \Sigma^{\lambda}_{i,(\tau)}  \mathbb{E} \left( \bm f_{t,(\tau)}\right)' \right)}{2\kappa_{2,(\tau)}^2}  \\
& - \kappa_{1,(\tau)}\frac{  X_{it} - \mathbb{E}\left( \bm \lambda_{i,(\tau)}\right) \mathbb{E} \left( \bm f_{t,(\tau)} \right)   }{\kappa_{2,(\tau)}^2}   + \left( 1+ \frac{\kappa_{1,(\tau)}^2}{2\kappa_{2,(\tau)}^2} \right) \mathbb{E} \left( z_{it,(\tau)}\right) \vphantom{\int_1^2} \Biggr]
\end{align*}
\State 5) Update $\bm f_{t,(\tau)}$ for $t=1,....,T$ from
\begin{equation}
q(\bm f_{t,(\tau)}  \vert \mathbf{x}) = N_{r}(\bm \mu^{f}_{t,(\tau)}, \bm \Sigma^{f}_{t,(\tau)}),
\end{equation}
where $ \bm \Sigma^{f}_{t,(\tau)} = \left( \frac{1}{\kappa_{2,(\tau)}^{2}}\sum_{i=1}^{n}  \mathbb{E} \left(\bm \lambda_{i,(\tau)}\right)  \mathbb{E} \left(\bm \lambda_{i,(\tau)}\right)^{\prime} \mathbb{E} \left(\frac{1}{\bm z_{it,(\tau)}} \right) \mathbb{E} \left( \frac{1}{ \sigma_{i,(\tau)}} \right) + \bm I_r \right)^{-1}$ and $\bm \mu^{f}_{t,(\tau)} = \bm  \Sigma^{f}_{t,(\tau)}\left(\frac{1}{\kappa_{2,(\tau)}^{2}} \sum_{i=1}^{n}  \mathbb{E} \left(\bm \lambda_{i,(\tau)}\right) x_{it} \mathbb{E}\left(\frac{1}{z_{it,(\tau)}} \right) \mathbb{E} \left( \frac{1}{\sigma_{i,(\tau)}} \right) - \right. \left. \frac{\kappa_{1,(\tau)}}{\kappa_{2,(\tau)}^{2}} \sum_{i=1}^n  \mathbb{E} \left( \frac{1}{\sigma_{i,(\tau)}} \right) \mathbb{E} \left(\bm \lambda_{i,(\tau)}\right)\right).$
\State r = r+1,
\EndWhile  
\end{algorithmic}}
\end{algorithm}
\doublespacing

It is evident from the formulas above that each parameter is dependent on the variational expectation of other parameters, such that update of all parameters cannot be achieved in one step. However, parameter updates can be achieved sequentially, similar to the EM-algorithm that is used to maximize likelihood functions. As with all iterative algorithms, it is important to ask whether there are convergence guarantees, whether a global maximum of the ELBO can be achieved, and how many iterations a typical run would require. \cite{Bleietal2017}, who provide an excellent introduction to these issues in general settings, note that different initialization of parameters will lead to slightly different paths of the ELBO. Therefore, as is the case with all EM-type algorithms, achieving a global optimum is not always guaranteed. Additionally, in many cases the ELBO might not converge to a fixed point, despite the fact that parameter updates from one iteration to the next are minimal. Therefore, it is important to choose good initial conditions. For the quantile factor model we initialize the factors to their PCA estimate, such that in step 1 of the algorithm we have during the first iteration $\mathbb{E} \left( \bm f_{t,(\tau)} \right) = \widehat{\bm f}_{t}^{pca}$. All other parameters are initialized to default values, i.e. they are either vectors of zeros or ones (for mean parameters) or set to $10\bm I$ (for variances/covariances).

\end{appendix}
 
\end{document}